\documentclass[lettersize,journal]{IEEEtran}
\usepackage{amsmath,amsfonts}

\usepackage{graphics} 
\usepackage{subfig}
\usepackage{epsfig} 
\usepackage{times} 
\usepackage{amsmath} 
\usepackage{amssymb}  
\usepackage{bm}
\usepackage{cite}
\usepackage{threeparttable}
\usepackage{multirow} 
\usepackage{booktabs}
\usepackage{colortbl}
\usepackage{xcolor}
\usepackage{algorithm}
\makeatletter
\renewcommand{\ALG@name}{\textit{Algorithm}}  
\makeatother
\usepackage{algpseudocode}

\usepackage{titlesec}
\titlespacing*{\subsection}{0pt}{6pt}{3pt}
\titlespacing*{\subsubsection}{0pt}{5pt}{2pt}
\allowdisplaybreaks

\begin{document}
\title{Dual-Interaction-Aware Cooperative Control Strategy for Alleviating Mixed Traffic Congestion}
	
\author{Zhengxuan Liu,~\IEEEmembership{Student~Member,~IEEE,} Yuxin Cai,~\IEEEmembership{Student~Member,~IEEE,} Yijing Wang, Xiangkun He,~\IEEEmembership{Senior~Member,~IEEE,} Chen Lv,~\IEEEmembership{Senior~Member,~IEEE} and Zhiqiang Zuo,~\IEEEmembership{Senior~Member,~IEEE} 
\thanks{*This work was supported by the National Natural Science Foundation of China (grants 62173243 and 61933014). }
\thanks{
Zhengxuan Liu, Yijing Wang and Zhiqiang Zuo are with
the Tianjin Key Laboratory of Intelligent Unmanned Swarm Technology and
System, School of Electrical and Information Engineering, Tianjin University,
Tianjin 300072, China. (e-mail: liuzhengxuan@tju.edu.cn; yjwang@tju.edu.cn; zqzuo@tju.edu.cn)}
\thanks{
Yuxin Cai, and Chen Lv are with School of Mechanical and Aerospace Engineering, 
Nanyang Technological University, Singapore. (e-mail: caiy0039@e.ntu.edu.sg; lyuchen@ntu.edu.sg)}
\thanks{
Xiangkun He is with Shenzhen Institute for Advanced Study, University of Electronic Science and Technology of China. (e-mail: xiangkun.he@uestc.edu.cn)}
}

\markboth{Journal of \LaTeX\ Class Files,~Vol.~14, No.~8, August~2021}%
{Shell \MakeLowercase{\textit{et al.}}: A Sample Article Using IEEEtran.cls for IEEE Journals}


\maketitle

\newcommand{\textbfmathcal}[1]{\bm{\mathcal{#1}}}

\begin{abstract}
As Intelligent Transportation System (ITS) develops, Connected and Automated Vehicles (CAVs) are expected to significantly reduce traffic congestion
through cooperative strategies, such as in bottleneck areas.
However, the uncertainty and diversity in the behaviors of Human-Driven Vehicles (HDVs) in mixed traffic environments present major challenges for CAV cooperation.
\textcolor{black}{This paper proposes a Dual-Interaction-Aware Cooperative Control (DIACC) strategy that enhances both local and global interaction perception within the Multi-Agent Reinforcement Learning (MARL) framework for Connected and Automated Vehicles (CAVs) in mixed traffic bottleneck scenarios.
The DIACC strategy consists of three key innovations:
1) A Decentralized Interaction-Adaptive Decision-Making (D-IADM) module that enhances actor's local interaction perception by distinguishing CAV-CAV cooperative interactions from CAV-HDV observational interactions.
2) A Centralized Interaction-Enhanced Critic (C-IEC) that improves critic's global traffic understanding through interaction-aware value estimation, providing more accurate guidance for policy updates.
3) A reward design that employs softmin aggregation with temperature annealing to prioritize interaction-intensive scenarios in mixed traffic.
Additionally, a lightweight Proactive Safety-based Action Refinement (PSAR) module applies rule-based corrections to accelerate training convergence.
Experimental results demonstrate that DIACC significantly improves traffic efficiency and adaptability compared to rule-based and benchmark MARL models.}
\end{abstract}
        

\vspace{-8pt}

\section{Introduction}
\IEEEPARstart{W}{ith} the rapid growth of vehicle ownership, traffic congestion has become a critical challenge, 
and connected and automated vehicles (CAVs) have demonstrated significant potential for improving traffic performance \cite{2022-TITS-AlexanderKatriniok, he2023fear}.
However, Human-Driven Vehicles (HDVs) will continue to coexist with CAVs for the foreseeable future \cite{xu2024multi}.
Human drivers exhibit diverse and unpredictable driving styles, posing a fundamental challenge for CAVs that must adapt to uncertain HDV behaviors while coordinating toward overall traffic optimization.

\begin{figure}[ht]
\centering
\subfloat[In purely HDV traffic, human drivers typically prioritize their own driving efficiency, 
leading to non-cooperative actions and congestion in the merging area. Different colors represent HDVs with various driving styles.]{
        \includegraphics[width=0.4\textwidth]{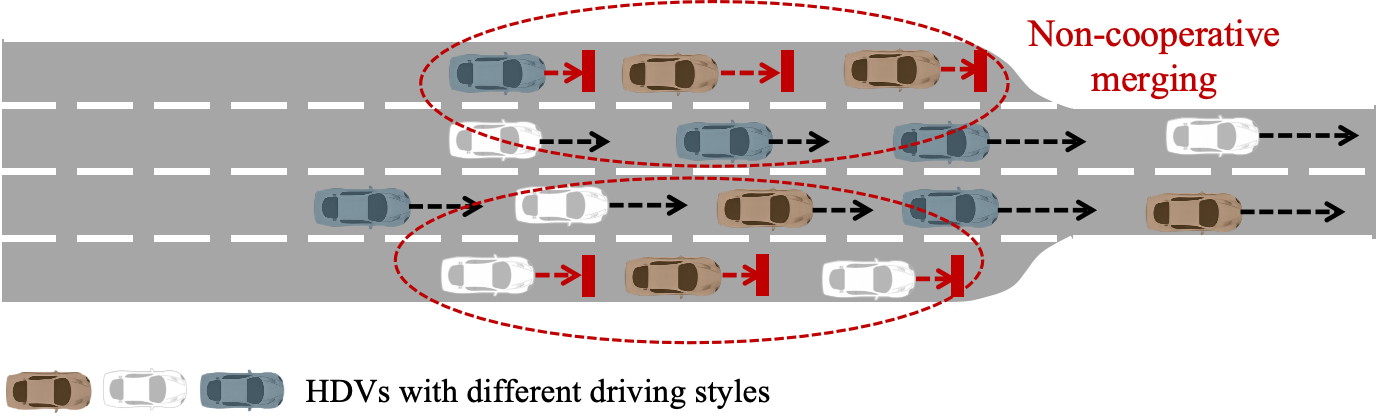}
        \label{fig:scenario_NC}
}
\hfill
\subfloat[In mixed traffic, CAVs engage in cooperative merging behaviors, providing opportunities for neighboring vehicles to change lanes, 
thus alleviating congestion. Blue vehicles represent CAVs.]{
        \includegraphics[width=0.4\textwidth]{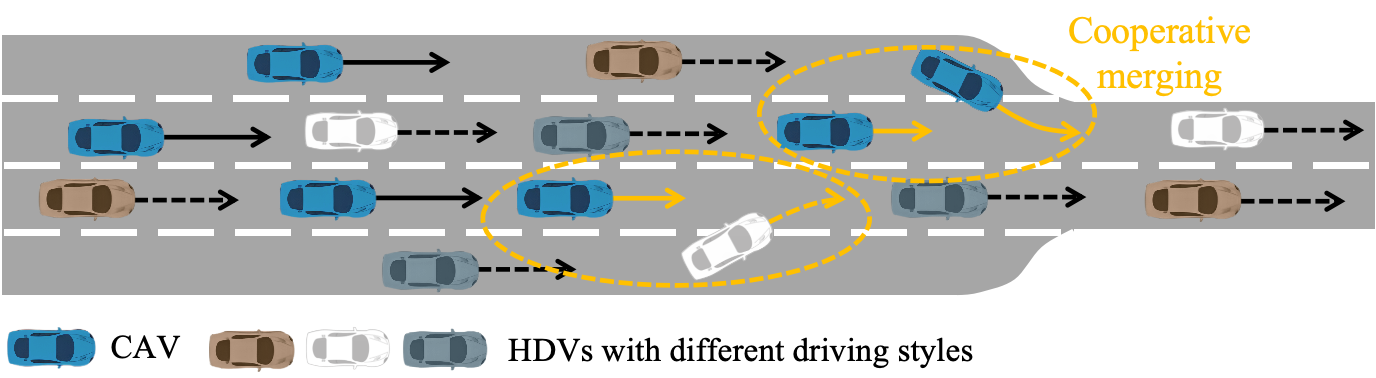}
        \label{fig:scenario_C}
}
\caption{Illustrations of non-cooperative merging in purely HDV traffic and cooperative merging with CAVs in a bottleneck scenario.}
\label{fig:scenario}
\end{figure}

This paper focuses on bottleneck merging scenarios characterized by high-density mixed traffic, where HDVs with varying driving styles coexist with cooperatively controlled CAVs (as depicted in Fig.~\ref{fig:scenario}).
Vehicle interaction is a critical determinant of traffic performance in bottleneck scenarios.
In purely HDV traffic, the lack of cooperation leads to severe congestion at the merging point, while CAVs with cooperative strategies can plan merging routes in advance and yield to surrounding vehicles to enhance overall efficiency.
However, the uncertainty and diversity of HDV behaviors significantly complicate the modeling of vehicle interactions, posing a core challenge for cooperative CAV control in mixed traffic.

For cooperative control of CAVs in bottleneck merging scenarios,
existing solutions fall into two categories.
The first includes rule-based or optimization-based approaches
\cite{chen2021harnessing, vcivcic2022coordinating, li2023developing, du2023adaptive, ding2023mpc},
while the second relies on Reinforcement Learning (RL)
\cite{wang2022integrated, li2022enhancing, chen2023deep, li2024nash}.
The centralized optimization method has advantages in addressing global traffic issues, but its computational cost is relatively high \cite{hu2019trajectory}. 
Furthermore, these methods often struggle to cope with complex traffic environments and vehicle interactions, 
as they heavily rely on the models of the traffic system and vehicle motion \cite{vcivcic2022coordinating, li2023developing, du2023adaptive}.
In contrast, RL-based methods are usually model-free, learning effective policies through interaction with the environment.
They have advantages in handling the uncertainties and complexities of mixed traffic environments. 
However, single-agent RL algorithms have limitations in designing effective cooperative strategies for CAVs in mixed traffic.
They are more suitable for solving local control problems \cite{wang2022integrated}.
Multi-Agent Reinforcement Learning (MARL) offers a more robust solution by facilitating coordination among multiple agents through interaction 
and information sharing, thereby improving the learning of cooperative strategies and enhancing training efficiency. 
Consequently, MARL is an appropriate and effective scheme for addressing the cooperative control of CAVs in the bottleneck scenarios discussed in this paper. 
However, in the decentralized MARL framework, cooperation between CAVs is restricted to local observations, limiting the improvement of global traffic efficiency.
In fact, the influence of vehicle interactions on traffic performance is a global problem, while existing MARL methods focus on local vehicle interactions and lack consideration for global traffic dynamics.

\textcolor{black}{As discussed above, vehicle interaction is a critical factor influencing traffic performance in bottleneck scenarios, 
however, it remains inadequately addressed in existing approaches.
To bridge this gap, we propose a Dual-Interaction-Aware Cooperative Control (DIACC) strategy that enhances interaction perception from both local and global perspectives within the MARL framework.}
The main contributions of this paper are as follows:
\begin{itemize}
\item \textcolor{black}{
To address the actor's inability to distinguish between heterogeneous vehicle interactions (i.e., CAVs, HDVs), a Decentralized Interaction-Adaptive Decision-Making (D-IADM) module is proposed.
The core design employs separate graph attention networks to differentiate CAV-CAV cooperative interactions (incorporating previous-timestep decision feedback) 
from CAV-HDV observational interactions that rely solely on historical trajectories.
Experimental results demonstrate improved training stability, higher traffic efficiency, and reduced safety-critical events.}
\item \textcolor{black}{
Centralized Interaction-Enhanced Critic (C-IEC) is devised to enable the critic to capture how vehicle interactions influence global traffic dynamics during training.
This provides a more accurate value estimation that guides D-IADM toward cooperative strategies accounting for system-level traffic dynamics.
The performance advantage is particularly pronounced in high-density scenarios.}
\item \textcolor{black}{
A cooperative reward mechanism with softmin aggregation and temperature annealing is presented.
This mechanism transitions from uniform exploration to optimization on difficult cases, thereby concentrating training on challenging interaction scenarios.
The resulting enhancement in training stability and convergence enables D-IADM to develop more robust cooperative strategies across diverse interactive scenarios.
}
\end{itemize}

The structure of the rest of this paper is as follows. 
Section II reviews the latest CAV cooperative control strategies.
Section III provides the fundamental mathematical descriptions for this study.
Section IV presents the framework of DIACC strategy and details the design of each module.
The experiment design and the corresponding analysis are conducted in Section V as well as the discussion of the findings. 
Section VI concludes this work.

To facilitate the explanation of the collaborative control strategy designed in this paper, Table \ref{tab:abbreviations} lists two categories of abbreviations: 
terminology abbreviations defined in our strategy and performance metric abbreviations used in the subsequent experimental evaluation.
\begin{table}[htpb]
\centering
\caption{Abbreviations.}
\fontsize{7pt}{9pt}\selectfont
\begin{tabular}{cc}
\hline
Full Name & Abbreviation \\
\hline
{\color{black}Centralized Training and Decentralized Execution} & {\color{black}CTDE}\\
Dual-Interaction-Aware Cooperative Control & DIACC\\
Decentralized Interaction-Adaptive Decision-Making & D-IADM\\
Trajectory-Aware Interaction Encoder & TAIE\\
\textcolor{black}{Proactive Safety-Based Action Refinement} & \textcolor{black}{PSAR}\\
Centralized Interaction-Enhanced Critic & C-IEC\\
Integrated Traffic Dynamics Representation & ITDR\\
MAPPO Model With D-IADM Module & MAPPO-IADM\\
\textcolor{black}{Safety-Critical Event} & \textcolor{black}{SCE}\\
\textcolor{black}{Waiting Event} & \textcolor{black}{WE}\\
\hline
\end{tabular}
\label{tab:abbreviations}
\end{table}
      
\vspace{-8pt}

\section{RELATED WORK}

Researchers have long focused on the issue of cooperative control of CAVs. 
The present results indicate that CAV cooperative control can significantly enhance traffic performance, such as traffic safety, efficiency, 
and energy consumption reduction.
The latest CAV cooperative control strategies are summarized in this section, including control theory-based approach, 
single-agent RL-based approach and MARL-based approach.
We analyze their characteristics, and thus clarify their advantages and limitations.
\subsection{Control Theory-based Approach}

The control theory-based approach typically relies on traffic flow theory 
(such as macroscopic and microscopic traffic flow models) \cite{vcivcic2022coordinating, ying2024infrastructure} 
and optimization algorithms, for example, Model Predictive Control (MPC) \cite{ding2023mpc, huang2024distributed}. 
When considering optimization objectives at the local or microscopic level, the focus is on car-following speed and lane-changing gap 
\cite{hou2023cooperative, chen2021harnessing}, and optimizing the vehicle trajectory according to these two factors. 
While at the global or macroscopic level, the main points of interest are typically the average speed and flow of the traffic \cite{vcivcic2022coordinating}. 
In these studies, vehicle behavior is guided by manually designed driving rules or objective functions \cite{shi2023cooperative, huang2024distributed}. 
The centralized optimization strategy can provide the optimal solution for global traffic issues like improving traffic efficiency but at high computational cost \cite{jing2019cooperative, ding2019rule}. 
Another significant limitation of control theory-based approach is the difficulty in handling complex traffic environments and vehicle interactions, 
as it usually depends on the modeling of traffic environment and vehicle behavior \cite{li2024dcoma, sun2024multi}.
However, it is challenging to model complex traffic evolution and vehicle interactions accurately.

\subsection{Single-Agent RL-based Approach}

The RL-based scheme has the advantage in dealing with environmental uncertainty and complexity. 
It does not rely on precise environmental modeling, but learns optimal decision strategies through interaction with the environment \cite{li2022enhancing}. 
Reinforcement learning guides the learning process by designing reward functions and training loss to achieve desired goals \cite{zhang2023learning, chen2023deep}. 
In cooperative control, researchers focus on two key issues: understanding interactions with vehicles in the traffic environment 
and enhancing cooperation among vehicles via communication. 
A Multi-View Graph Convolutional Reinforcement Learning (MVGRL) algorithm was proposed in \cite{xu2024multi}, which
addresses on the interactions among traffic participants and the resulting complex traffic states. 
\cite{li2024augmented} put forward the parameter-sharing mechanism and dense communication topology to enable CAVs to tackle 
traffic fluctuations and improve driving efficiency. 
However, it is worth noting that the single-agent reinforcement learning approach still has some drawbacks when dealing with the collaboration among CAVs in mixed traffic. 
On the one hand, single-agent approach is more suitable for solving local control problems \cite{wang2022integrated}, 
but it faces challenges in global traffic optimization, with the requirement of a large amount of interaction experience to learn effective decision strategies. 
This leads to high training cost and the difficulty of large-scale deployment. 
On the other hand, it is challenging for such an approach to handle coordination problems among multiple vehicles, 
often requiring additional rule guidance or manually designed coordination interaction settings. 
For example, \cite{bouton2019cooperation} evaluated the level of cooperation among drivers by learning vehicle interactions 
and used this cooperation belief state as the input to the RL algorithm.

\subsection{MARL-based Approach}

MARL helps to model and analyze interactions among agents in traffic environments \cite{irshayyid2024review}, 
and its own flexible cooperation and sharing mechanism also promote research on collaborative control.
Currently, the MARL approach has been applied to various traffic optimization problems, such as traffic light control \cite{chu2019multi, wang2024traffic}, 
variable speed limit control \cite{wang2022integrated}, multi-vehicle cooperative control \cite{spatharis2024multiagent, hu2019interaction}, 
and vehicle-road cooperative control \cite{yang2020urban, huang2024cooperative}. 
The advantages of MARL have been demonstrated in several aspects. 
First, it facilitates the cooperation through interaction and information sharing among multiple agents \cite{chen2023deep, li2022enhancing}, 
allowing the learning of optimal collaborative strategies.
Second, MARL can handle non-stationarity in the environment, especially as the number of agents increases \cite{madridano2021trajectory, bautista2022autonomous}.
Third, MARL shows its merits in terms of convergence, optimality, and stability \cite{yu2020distributed}.

In traffic systems, MARL offers new possibilities for achieving highly intelligent and adaptive traffic management. 
However, due to the complexity of interactions among multiple agents, 
the design of MARL algorithms for multi-vehicle cooperative control is in the early phase. 
For multi-vehicle collaboration, \cite{nakka2022multi} employed a multi-agent deep deterministic policy gradient algorithm to train several CAVs from the mainline and ramp, 
while \cite{chen2023deep} proposed a multi-agent advantage actor-critic algorithm to train all CAVs within a merging area. 
Additionally, cooperative lane-changing actions for vehicles were designed in \cite{li2022enhancing} through communication collaboration and parameter sharing. 
The purpose of these studies is mainly the design of individual agents in MARL algorithms, but the guiding role for critics has not been fully explored.
At the same time, they ignore the global impact of vehicle interaction on traffic optimization.

Within the framework of MARL involving CTDE, this paper highlights the important influence of vehicle interaction 
integrating both local and global perspectives.
Based on local interaction awareness, we design the D-IADM to enhance CAV adaptability to HDVs with unpredictable and diverse behaviors. 
To realize better guidance of vehicle cooperation, the C-IEC is suggested to investigate the effects of vehicle interactions on traffic efficiency 
with the help of global interaction awareness.
By doing so, we achieve a significant improvement in traffic efficiency in bottleneck scenarios and 
enhance the adaptability of the MARL model to various scenarios.

\vspace{-8pt}

\section{PROBLEM FORMULATION}
\subsection{Modeling Cooperative Decision-Making Problem}
In this study, we model the cooperative decision-making process of CAVs in bottleneck merging scenarios within a mixed traffic environment as an MARL problem, 
which is characterized by a decentralized partially observable Markov decision process (Dec-POMDP) \cite{Dec-POMDP_2016}. 
The MARL model can be represented as a tuple $(\mathcal{N}, \mathcal{S}, \textbfmathcal{A}, \textbfmathcal{O}, R, P, \gamma)$, where:
\begin{itemize}
    \item $\mathcal{N}=\{1,\ldots,n\}$ denotes the set of CAVs in the mixed traffic environment.
    \item $\mathcal{S}$ stands for the global state space of the environment.
    \item $\textbfmathcal{A}=\prod_{i=1}^{n} \mathcal{A}^{i}$ is the joint action space of all CAVs.
    \item $\textbfmathcal{O}=\prod_{i=1}^{n} \mathcal{O}^{i}$ denotes the joint observation space of all CAVs. 
    \item $R: \textbfmathcal{O} \times \textbfmathcal{A} \rightarrow \mathbb{R}$ specifies the joint reward function.
    \item $P(s_{t+1}|s_t, a^1_t, \ldots, a^n_t) : \mathcal{S} \times \textbfmathcal{A} \times \mathcal{S} \rightarrow [0,1]$ 
    defines the state transition probability function.
    \item $\gamma \in [0,1)$ is the discount factor, balancing the importance of immediate and future rewards.
\end{itemize}
In this setup, each CAV $i$ at any time step $t$ observes its local state $o_{t}^{i} \in \mathcal{O}^{i}$ 
and selects an action $a_{t}^{i}$ according to its policy $\pi_{\theta}^{i}(\cdot | o_{t}^{i})$ parameterized by $\theta$.

The collective objective for all CAVs is to optimize the team policy $\pi_{\theta}^{*}$, aiming to maximize the expected cumulative reward 
over a horizon of $L$ time steps, that is,
\begin{equation}\label{eq:objective}
    J(\pi_{\theta}) = \mathbb{E}_{\pi_{\theta}}\left[\sum_{t=0}^{L} \gamma_{t} R_t\right].
\end{equation}

\subsection{Multi-Agent Actor-Critic Framework}
In this subsection, we utilize MARL with CTDE, which consists of two pivotal parts: a centralized critic network and $n$ decentralized actor networks. 
The former supplements the training phase by accessing global information from all agents, thus mitigating the non-stationary of the environment 
and enhancing learning efficiency.
Concurrently, each decentralized actor network strives to maximize its individual reward, benefiting from the global insights provided by the critic.
A well-known representative is MAPPO \cite{MAPPO}, which extends Proximal Policy Optimization (PPO) \cite{schulman2017proximal} to multi-agent setting.
MAPPO is characterized by its ability to accommodate the complexities of multiple interacting agents, making it particularly effective for developing 
decentralized control strategies where each agent independently interacts with the environment and other agents.

\color{black}
In MAPPO, each agent's actor network $\pi_\theta^i$ maps its local observation $o_{t}^{i}$ to a categorical distribution over the discrete action space $\mathcal{A}^{i}$, with the aim of optimizing its policy $\pi_\theta^i(a_t^i|o_t^i)$ to maximize the expected reward.
Specifically, the actor networks focus on optimizing the clipped surrogate objective \cite{schulman2017proximal},
\begin{equation}\label{eq:clipped_objective}
L^{CLIP}(\theta) = \mathbb{E}\left[\min\left(r_t^i(\theta) \hat{A}_t, \text{clip}(r_t^i(\theta), 1 - \epsilon, 1 + \epsilon) \hat{A}_t \right)\right]
\end{equation}
where $r_t^i(\theta) = \frac{\pi_{\theta}(a_t^i|o_t^i)}{\pi_{\theta_{\text{old}}}(a_t^i|o_t^i)}$
represents the probability ratio between the new and old policies.
The clipping mechanism governed by hyperparameter $\epsilon$ prevents the policy from deviating too far from the old policy, which ensures the updating stability.
In this objective, the advantage estimator $\hat{A}_t$ is derived according to Generalized Advantage Estimator (GAE),
which provides a bias-variance trade-off for accurate advantage estimation.
The joint advantage estimate is calculated by:
\begin{align}\label{eq:advantage}
    \hat{A}_t^{GAE} (\gamma, \lambda) = \sum_{l=0}^{T-t} (\gamma \lambda)^l \delta_{t+l},
\end{align}
where $\lambda$ is the GAE parameter and $T$ is the episode length.
\textcolor{black}{$\delta_{t}$ is the TD error with the form of}
\begin{equation}\label{eq:td_error}
    \textcolor{black}{\delta_{t} = R_t + \gamma V_{\phi}(s_{t+1}) - V_{\phi}(s_t),}
\end{equation}
\textcolor{black}{where $R_t$ is the immediate reward and $V_{\phi}$ is the centralized critic's value function (detailed below). 
The TD error quantifies the discrepancy between the actual return and the critic's estimate, which is accumulated via GAE to compute the advantage for policy updates.}

In addition to the clipped objective, an entropy term is incorporated into the actor's loss to prevent premature convergence and promote sufficient exploration:
\begin{equation}\label{eq:actor_loss}
L(\theta) = L^{CLIP}(\theta) + \sigma \mathbb{E}\left[ H(\pi_{\theta}(a_t^i|o_t^i))\right]
\end{equation}
where $H$ denotes the entropy of the policy, and $\sigma$ is the entropy coefficient.
The inclusion of entropy rewards facilitates exploration and avoids overfitting to specific scenario features, fostering action diversity and
enhances generalization capability in similar states.

While each actor independently optimizes its local policy based on individual observations, the centralized critic leverages global state information to guide the training process.
The centralized critic, denoted as $V_\phi$, performs the mapping: $\mathcal{S} \rightarrow \mathbb{R}$.
It provides a global value function $V_{\phi}(s_{t})$ that evaluates the current state $s_t$, guiding the policy update for each agent.
To stabilize the critic's updates, the centralized critic, parameterized by $\phi$, is also trained using clipped value predictions:
\begin{equation}\label{eq:clipped_value}
V^{CLIP}(\phi) = \text{clip}(V_{\phi}(s_t), V_{\phi_{old}}(s_t) - \epsilon, V_{\phi_{old}}(s_t) + \epsilon)
\end{equation}
The value function loss {\color{black}has} the form
\begin{equation}\label{eq:value_loss}
L_{V}(\phi) = \mathbb{E}\left[ \max\left((V_{\phi}(s_t) - \hat{R}_t)^2, (V^{CLIP}(\phi) - \hat{R}_t)^2\right)\right]
\end{equation}
where $\hat{R}_t$ represents the discounted reward-to-go $\sum_{t' = t}^{T} \gamma_{t'} R(s_{t'}, a_{t'}, s_{t'+1})$, 
indicating the expected future return.

To summarize, we adopt the MAPPO framework to learn decentralized policies ($\pi_{\theta}^{1},\pi_{\theta}^{2}, ..., \pi_{\theta}^{n}$)
where each agent's independent policy update is clipped in terms of the objective (\ref{eq:clipped_objective}).
Then, a centralized critic is employed to estimate the value function, which provides shared information across agents to stabilize learning and improve coordination.
\textcolor{black}{However, applying vanilla MAPPO to mixed traffic cooperative control reveals several limitations, which motivate the DIACC strategy presented in the next section:
\begin{itemize}
\item The actor does not explicitly model the interactions among agents, which is particularly challenging in mixed traffic where CAV-CAV cooperative interactions and CAV-HDV observational interactions exhibit fundamentally different characteristics. To address this, we propose the D-IADM module, which focuses on modeling the local interaction relationships of CAVs in mixed traffic environments.
\item Although the centralized critic has access to global information, it lacks effective mechanisms to fully exploit such information, 
making it difficult to obtain accurate value estimation. To this end, we propose C-IEC, which leverages the relationship mapping between global vehicle interactions and traffic dynamics to provide more accurate value estimates for guiding policy updates.
\item The standard reward design does not adequately emphasize interaction-intensive scenarios where agents face complex coordination challenges, 
potentially neglecting critical situations in bottleneck merging. To overcome this, we propose a cooperative reward mechanism with softmin aggregation and temperature annealing, which directs training attention toward difficult interaction scenarios and enhances learning robustness.
\end{itemize}}

\color{black}
\begin{figure*}[htpb]
\centering
\includegraphics[width=1.0\textwidth]{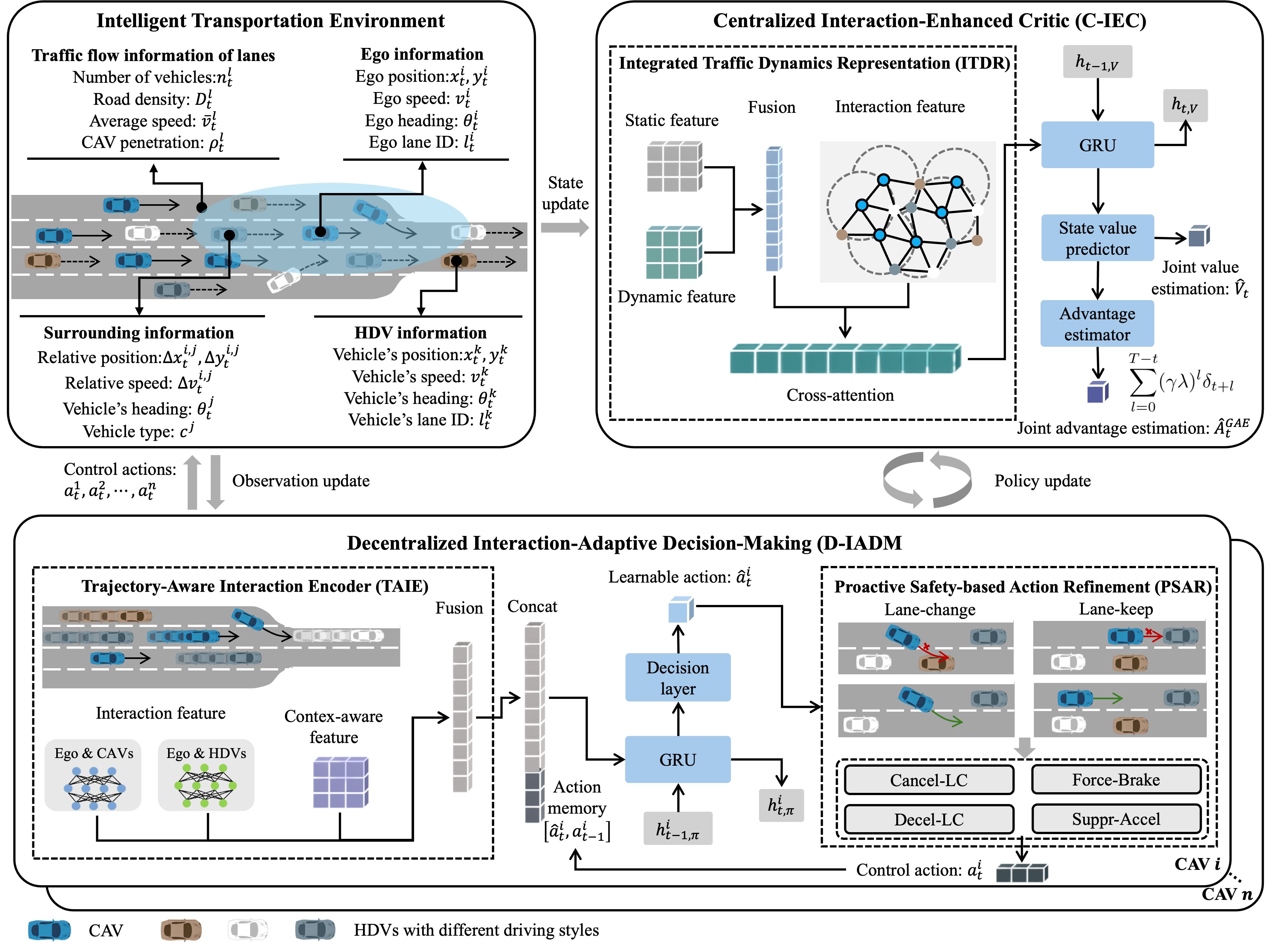}
\caption{Framework of DIACC strategy.}
\label{fig:framework}
\end{figure*}

\vspace{-8pt}

\section{Dual-Interaction-Aware Cooperative Control Strategy for CAVs}
With the MAPPO framework, we propose Dual-Interaction-Aware Cooperative Control (DIACC) strategy for CAVs,
to enhance the effectiveness of cooperative control in bottleneck scenarios.
This section details the design and implementation of this strategy.

\vspace{-8pt}
\subsection{Framework}
The proposed cooperative control model combines a multi-agent actor-critic framework, including several {\color{black}Decentralized} Interaction-Adaptive Decision-Making (D-IADM) modules
and a Centralized Interaction-Enhanced Critic (C-IEC), as shown in Fig. \ref{fig:framework}. 
During the training process, each CAV acquires local observation from the environment and outputs its control action through the D-IADM. 
The D-IADM first obtains interaction features between the ego and surrounding vehicles via a Trajectory-Aware Interaction Encoder (TAIE). 
Then, it combines the last decision output $a_{t-1}^i$ with a Gated Recurrent Unit (GRU) \cite{cho2014learning} 
and a decision layer to generate the learnable action output $\hat{a}_{t}^i$. 
The GRU captures temporal dependencies between observation features across different time steps, enhancing the continuity of decision-making. 
\textcolor{black}{The output $\hat{a}_{t}^i$ is subsequently fed into the PSAR module, a lightweight action refinement layer that monitors spacing and TTC during lane-changing and car-following maneuvers. 
When the spacing or TTC falls below a safety threshold, PSAR applies minor corrections to $\hat{a}_{t}^i$ to mitigate potential collision risks.}
The C-IEC module receives all state information from the environment and outputs a global state value function to assess the overall state quality 
and guide parameter updating in the D-IADM. 
In order to better understand the evolution of traffic and guide the movement of CAVs, we use Integrated Traffic Dynamics Representation (ITDR) module to 
capture the key features of traffic dynamics and interactions among vehicles. 

The details of state space, action space and reward function are illustrated in the following.

\vspace{-8pt}
\subsubsection{State Space}

The vehicular states consist of the position $(x,y)$, speed $v$, heading angle $\theta$, lane index $l$, and vehicle type $c$. 
Furthermore, lane-specific information, including the number of vehicles $n^l$, density $D^l$, average speed $\overline{v}^l$ is also considered, 
and CAV penetration rate $\rho^l$, where $l$ represents the index of each lane in the bottleneck scenario. 
Additionally, detailed road structure information is provided by high-precision maps. 
These data form the state information that the centralized critic can get from the environment.

To help CAVs better understand interaction features with their surroundings, we design local observation states for each CAV $i$, more specifically, 
\begin{itemize}
\item \textbf{Self-State Information}: Position coordinates $(x^{i}_t, y^{i}_t)$, speed $v^{i}_t$, heading angle $\theta^{i}_t$, and lane index $l^{i}_t$.
\item \textbf{Neighboring Vehicle State Information}: This is defined by a set $\mathcal{M}_{s}^{i}$, which includes six closest vehicles in the same and adjacent lanes. 
The state of each neighboring vehicle $j \in \mathcal{M}_{s}^{i}$ relative to CAV $i$ is described by relative position $(\Delta x_{t}^{i,j}, \Delta y_{t}^{i,j})$ and 
relative speed $\Delta v_{t}^{i,j}$ as well as heading angle $\theta^{j}_t$ and vehicle type $c^{j}$ of neighboring vehicles.
\item \textbf{Neighboring Lane Traffic Flow Statistics}: CAVs understand surrounding traffic status through shared lane statistics, 
including the number of vehicles $n^l$, density $D^l$, average speed $\overline{v}^l$, and CAV penetration rate $\rho^l$.
\item \textbf{Static Road Structure Information}: 
The coordinates and lane counts of sections before and after the merging point
are used as static road structure information.
\end{itemize}

\vspace{-8pt}
\subsubsection{Action Space}
Each CAV $i$ has a discrete action space $\mathcal{A}_i$, encompassing lateral control and longitudinal control. 
Specifically, $\mathcal{A}_i$ contains five decision behaviors: maintain current lane and speed (denoted as ``remain"), lane change to the left or right, accelerate, or decelerate. 
Thus, the action set for each CAV is formally defined as $\mathcal{A}_i = \{0: \text{remain}, 1: \text{left}, 2: \text{right}, 3: \text{accelerate}, 4: \text{decelerate}\}$. 
At each time step, CAV $i$ selects an action $a_{t}^{i} \in \mathcal{A}_i$ through its policy $\pi_{\theta}^{i}$, to maximize its expected cumulative reward.

\subsubsection{Reward Function}
\color{black}
The total reward comprises three components: a softmin-aggregated local ego reward $\bar{r}_{e}$, a global reward $r_{g}$, and a completion reward $r_{done}$:
\begin{equation}\label{eq:reward_all}
    R_t = w_{e} \bar{r}_{e} + w_{g} r_{g} + r_{done},
\end{equation}
where $w_e$ and $w_g$ are fixed weighting coefficients.  
The reward $r_{done}$ is also included to incentivize CAVs to traverse the bottleneck area, thereby improving throughput.

In mixed traffic bottleneck scenarios, CAVs encounter heterogeneous interaction complexity: 
some operate in sparse regions with simple dynamics, while others navigate dense merging areas with intensive vehicle interactions, constituting the interaction-intensive cases.
To direct learning toward these cases, a softmin-based aggregation is applied to the local ego rewards:
\begin{equation}\label{eq:softmin_reward}
\bar{r}_{e} = \sum_{i=1}^{n} w^{i} r_{e}^{i},
\end{equation}
where the softmin weights are computed as:
\begin{equation}\label{eq:softmin_weight}
w^{i} = \frac{\exp(-r_{e}^{i} / \tau)}{\sum_{j=1}^{n} \exp(-r_{e}^{j} / \tau)}.
\end{equation}
The weight $w^{i} \propto \exp(-r_e^i/\tau)$ assigns higher importance to agents with lower rewards, 
thereby directing the learning focus toward poorly-performing agents and the most challenging cooperative scenarios at the bottleneck.
The temperature parameter $\tau$ controls the focus intensity: a large $\tau$ yields nearly uniform weighting for broad exploration, while a small $\tau$ concentrates learning on the worst-performing agents.
To leverage both regimes throughout training, a linear annealing schedule is employed:
\begin{equation}\label{eq:tau_annealing}
    \tau(t) = \begin{cases}
        \tau_{init} - \frac{(\tau_{init} - \tau_{final}) \cdot t}{T_{anneal}} & \text{if } t \leq T_{anneal}, \\
        \tau_{final} & \text{if } t > T_{anneal},
    \end{cases}
\end{equation}
where $\tau_{init}$ and $\tau_{final}$ are the initial and final temperature values, and $T_{anneal}$ is the annealing duration.
This curriculum-inspired strategy \cite{bengio2009curriculum} enables broad exploration in early training and focused optimization on difficult interactions in later stages, preventing the policy from being dominated by agents in simpler environments.
The ego reward $r_{e}^{i}$ consists of several sub-reward components, that is,
\begin{align}
    r_{e}^{i} &= w_{e,v} r_{e,v}^{i} + w_{e,w} r_{e,w}^{i} + w_{e,c} r_{e,c}^{i} + w_{e,tp} r_{e,tp}^{i}, \nonumber \\
    r_{e,v}^{i} &= -\frac{|v^{i} - v_{max}|}{v_{max}} + p_{e,v}, \nonumber \\
    r_{e,w}^{i} &= \sum_{j \in \mathcal{M}_{w}^{i}} -\frac{d_{th,w} - \sqrt{{\Delta x^{i,j}}^2 + {\Delta y^{i,j}}^2}}{d_{th,w} - d_{th,c}}, \nonumber \\
    r_{e,c}^{i} &= \sum_{j \in \mathcal{M}_{c}^{i}} \left[-\frac{d_{th,c} - \sqrt{{\Delta x^{i,j}}^2 + {\Delta y^{i,j}}^2}}{d_{th,c}} + p_{e,c}\right], \nonumber \\
    r_{e,tp}^{i} &= \text{sigmoid}(v^i-v_{th}) - 1. \label{eq:reward_ego}
\end{align}
where $w_{e,v}$, $w_{e,w}$, $w_{e,c}$, \textcolor{black}{and $w_{e,tp}$} are weighting parameters for speed, warning distance, collision penalty, \textcolor{black}{and time penalty, respectively}.
The parameter $v_{max}$ represents the prescribed maximum speed.
The sets $\mathcal{M}_{w}^{i}$ and $\mathcal{M}_{c}^{i}$ denote vehicles within the warning and actual collision ranges, with distance thresholds $d_{th,w}$ and $d_{th,c}$, respectively.
The parameters $p_{e,v}$ and $p_{e,c}$ are used to adjust the speed reward and collision penalty.
\textcolor{black}{The time penalty $r_{e,tp}^{i}$ penalizes vehicles with speeds below threshold $v_{th}$ to prevent lazy behavior.}

The global reward $r_{g}$ reflects the overall performance of all CAVs in terms of average speed:
\begin{equation}\label{eq:reward_global}
    r_{g} = -\frac{\left|\frac{1}{n}\sum_{i=1}^{n} v^i - v_{max}\right|}{v_{max}} + p_{g,v}.
\end{equation}
where $p_{g,v}$ is a parameter to adjust the global speed reward.

\subsection{D-IADM Module}
\textcolor{black}{The D-IADM module enhances the actor's local interaction perception by distinguishing CAV-CAV cooperative interactions from CAV-HDV observational interactions.
This module consists of two key components: 
1) the Trajectory-Aware Interaction Encoder (TAIE) that models the two interaction types with separate attention networks, and 
2) the Proactive Safety-based Action Refinement (PSAR) module that applies rule-based corrections to reduce high-risk actions.
The action memory mechanism incorporates both the actor-generated action and the PSAR-refined action into the next time step's observation, 
providing the actor with a comprehensive record of both intended and executed decisions.
}

\color{black}
\subsubsection{Trajectory-Aware Interaction Encoder (TAIE)}
\begin{figure}[htpb]
\centering
\includegraphics[width=0.5\textwidth]{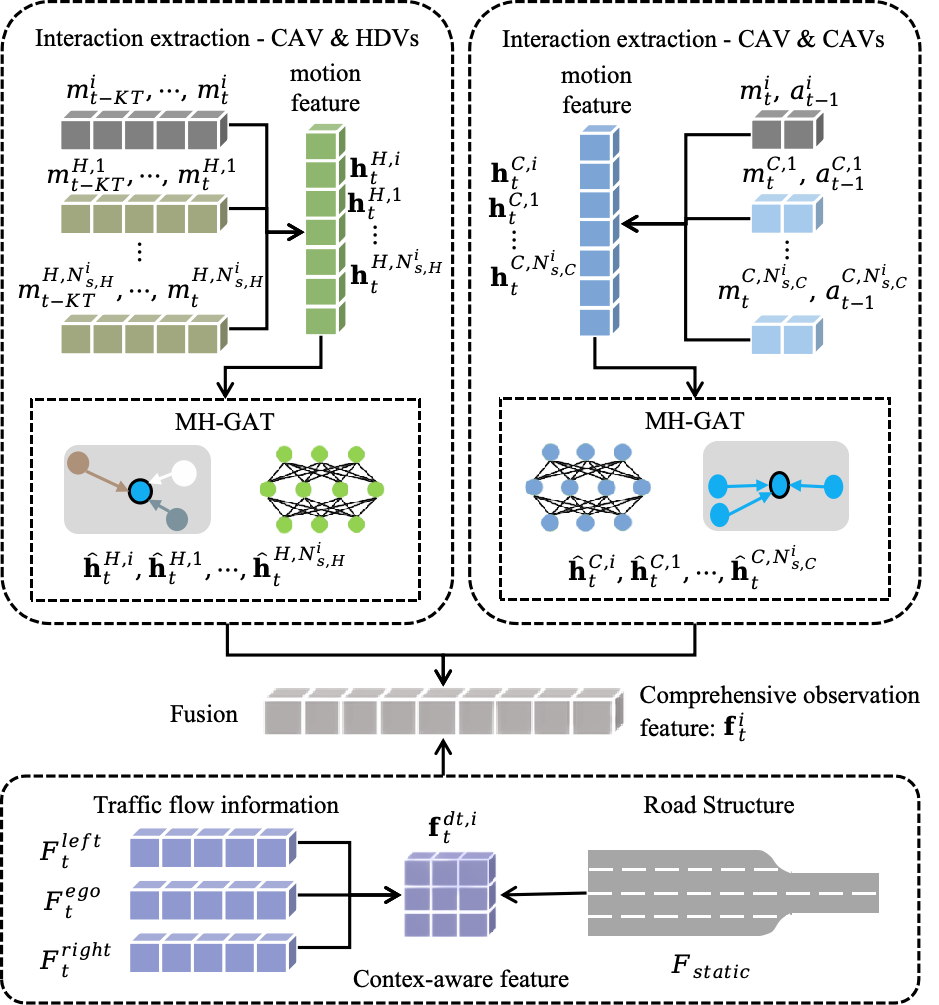}
\caption{Network framework of the TAIE.}
\label{fig:TIE}
\end{figure}

The purpose of TAIE module is to process each CAV's local observations and extract key interaction features necessary for decision-making, as illustrated in Fig. \ref{fig:TIE}. 
Each CAV's local observation state includes its own state information, neighboring vehicle motion information, lane statistics, and static road structure information. 
The set of neighboring vehicles, denoted as $\mathcal{M}_{s}^{i}$, contains six closest vehicles in the same and adjacent lanes. 
Considering the behavioral differences between HDVs and CAVs, we categorize neighboring vehicles into HDV sets $\mathcal{M}_{s,H}^{i}$ and CAV sets $\mathcal{M}_{s,C}^{i}$. 
This classification enhances the model's ability to capture interactions between CAVs and HDVs and reduces the bias.

To illustrate the interaction modeling process, we take the extraction of HDV interaction features as an example. 
To this end, an interaction graph between CAV $i$ and surrounding HDVs is constructed, which connects each CAV to its neighboring HDVs within a specified range. 
The HDV interaction graph is represented by $\mathcal{G}_{s,H}^{i} = (\mathcal{V}_{s,H}^{i}, \mathcal{E}_{s,H}^{i})$, where $\mathcal{V}_{s,H}^{i}=i \,\cup \mathcal{M}_{s,H}^{i}$ 
denotes the nodes and $\mathcal{E}_{s,H}^{i}$ stands for the edges connecting these nodes. 
Note that these edges only connect CAV $i$ with its neighboring HDVs, and there are no connections among inter-HDVs.
\textcolor{black}{
Denote $m_{t}^{i}=(x_{t}^{i}, y_{t}^{i}, v_{t}^{i}, \theta_{t}^{i})$ as the state of vehicle $i$,
where $x_{t}^{i}$, $y_{t}^{i}$, $v_{t}^{i}$, and $\theta_{t}^{i}$ represent position, speed, and heading angle, respectively.
The CAV-HDV layer input is composed of current and historical states of CAV $i$ and its $N_{s,H}^i$ surrounding HDVs.
The historical states are concatenated and embedded into node features $\mathbf{h}^{H, j}_{t} \in \mathbb{R}^{S_H}$,
which are then fed into a Multi-Head Graph Attention (MH-GAT) \cite{velivckovic2017graph} to extract interaction features.}

The enhanced node feature $\hat{\mathbf{h}}^{H, i}_{t}$ for CAV $i$ is obtained by aggregating adjacent features weighted across all $B$ attention heads:
\begin{equation}
\hat{\mathbf{h}}^{H, i}_{t} = \sigma \left( \frac{1}{B} \sum_{b=1}^{B} \sum_{j \in \mathcal{M}_{s,H}^{i}} \alpha_{ij}^b W^b \mathbf{h}^{H, j}_{t} \right),
\end{equation}
where $\alpha_{ij}$ is the attention weight quantifying the relative importance of surrounding HDV $j$ to CAV $i$, computed as:
\begin{equation}
\alpha_{ij} = \frac{\exp(\text{LeakyReLU}(\hat{\mathbf{a}} [W\mathbf{h}^{H, i}_t \| W\mathbf{h}^{H, j}_t]))}{\sum_{k \in \mathcal{M}_{s,H}^{i}} \exp(\text{LeakyReLU}(\hat{\mathbf{a}} [W\mathbf{h}^{H, i}_t \| W\mathbf{h}^{H, k}_t]))}
\end{equation}
with $W$ being the weight matrix and $\mathbf{\hat{a}}$ a single-layer feedforward network followed by LeakyReLU nonlinearity and softmax normalization.
This aggregation produces a rich representation of the HDV interaction dynamics around CAV $i$.

A similar process is performed to generate the CAV interaction graph. 
\textcolor{black}{Unlike the HDV graph, the CAV graph additionally incorporates the previous decision information $a_{t-1}^{i}$ of each CAV, yielding CAV interaction features $\mathbf{f}_t^{C,i}$.
This design reflects the key difference: CAV-CAV interactions involve shared decision information, while CAV-HDV interactions rely on observed trajectory patterns.}
\textcolor{black}{Additionally, local traffic context information (lane statistics $\mathbf{F}^{ego}_{t}$, $\mathbf{F}^{left}_{t}$, $\mathbf{F}^{right}_{t}$ and static road structure $\mathbf{F}_{static}$) is processed through an MLP to obtain the context embedding $\mathbf{f}_{t}^{dt, i}$.}
Finally, the HDV interaction embedding, CAV interaction embedding, and local traffic context embedding are combined in an MLP-based fusion layer, 
forming the comprehensive observation embedding feature $\mathbf{f}_{t}^{i}$ at time step $t$.

\subsubsection{\textcolor{black}{Proactive Safety-based Action Refinement (PSAR)}}

\color{black}
\begin{algorithm}[htpb]
\color{black}
\caption{\color{black}Proactive Safety-based Action Refinement (PSAR)}
\label{alg:PSAR}
\begin{algorithmic}[1]
\Require Actor action $\hat{a}_t^i \in \{0,1,2,3,4\}$; states of ego and surrounding vehicles; \\
  \hspace*{1.2em}Distance thresholds $d_\text{LC,min} \leq d_\text{safe} < d_\text{warn} < d_\text{att}$;\quad TTC thresholds $\tau_\text{safe} {<} \tau_\text{warn} {<} \tau_\text{att}$; \\
  \hspace*{1.2em}$\text{Risk}(d_0,\tau_0) \equiv |d_{t,long}^{i,k}| \leq d_0 \;\wedge\; 0 < \text{TTC}_t^{i,j} \leq \tau_0$ (for the relevant vehicle pair in the current context); 
\Ensure Refined action $a_t^i$
\State $a_t^i \leftarrow \hat{a}_t^i$
\If{$a_t^i \in \{1,2\}$}
    \State $b_{lc} \leftarrow 0$;
    \If{target-lane front vehicle $j$ exists}
        \If{$d_{t,long}^{i,j} \leq d_\text{LC,min}$ \textbf{or} $\text{Risk}(d_\text{safe},\tau_\text{safe})$}
            \State $a_t^i \leftarrow 0$ \Comment{Cancel-LC}
        \ElsIf{$\text{Risk}(d_\text{safe},\tau_\text{att})$ \textbf{or} $\text{Risk}(d_\text{att},\tau_\text{safe})$}
            \State $b_{lc} \leftarrow \min(|\Delta v_{t,long}^{i,j}|, b_\text{max})$ \Comment{Decel-LC}
        \EndIf
    \EndIf
    \If{target-lane rear vehicle $k$ exists \textbf{and} $a_t^i \neq 0$}
        \If{$|d_{t,long}^{i,k}| \leq d_\text{LC,min}$ \textbf{or} $\text{Risk}(d_\text{safe},\tau_\text{safe})$}
            \State $a_t^i \leftarrow 0$ \Comment{Cancel-LC}
        \ElsIf{$[\text{Risk}(d_\text{safe},\tau_\text{att})$ \textbf{or} $\text{Risk}(d_\text{att},\tau_\text{safe})]$ \textbf{and} $b_{lc} \neq 0$}
            \State $a_t^i \leftarrow 0$ \Comment{Cancel-LC}
        \EndIf
    \EndIf
\EndIf
\If{$a_t^i \in \{0,3,4\}$}
    \If{front vehicle $j$ exists} 
        \If{$d_{t,long}^{i,j} \leq d_\text{warn}$ \textbf{and} $\text{TTC}_t^{i,j} {>} 0$}
            \State $a_t^i \leftarrow 4$; \Comment{Force-Brake}
        \ElsIf{$d_{t,long}^{i,j} \leq d_\text{safe}$ \textbf{or} ($d_{t,long}^{i,j} \leq d_\text{warn}$ \textbf{and} $a_t^i {=} 3$)}
            \State $a_t^i \leftarrow 0$ \Comment{Suppr-Accel}
        \ElsIf{$\text{Risk}(d_\text{att},\tau_\text{warn})$}
            \State $a_t^i \leftarrow 4$; \Comment{Force-Brake}
        \ElsIf{$\text{Risk}(d_\text{att},\tau_\text{att})$ \textbf{and} $a_t^i {=} 3$}
            \State $a_t^i \leftarrow 0$ \Comment{Suppr-Accel}
        \EndIf
    \EndIf
\EndIf
\State \Return $a_t^i$
\end{algorithmic}
\end{algorithm}

\begin{figure}[htpb]
\centering
\includegraphics[width=0.25\textwidth]{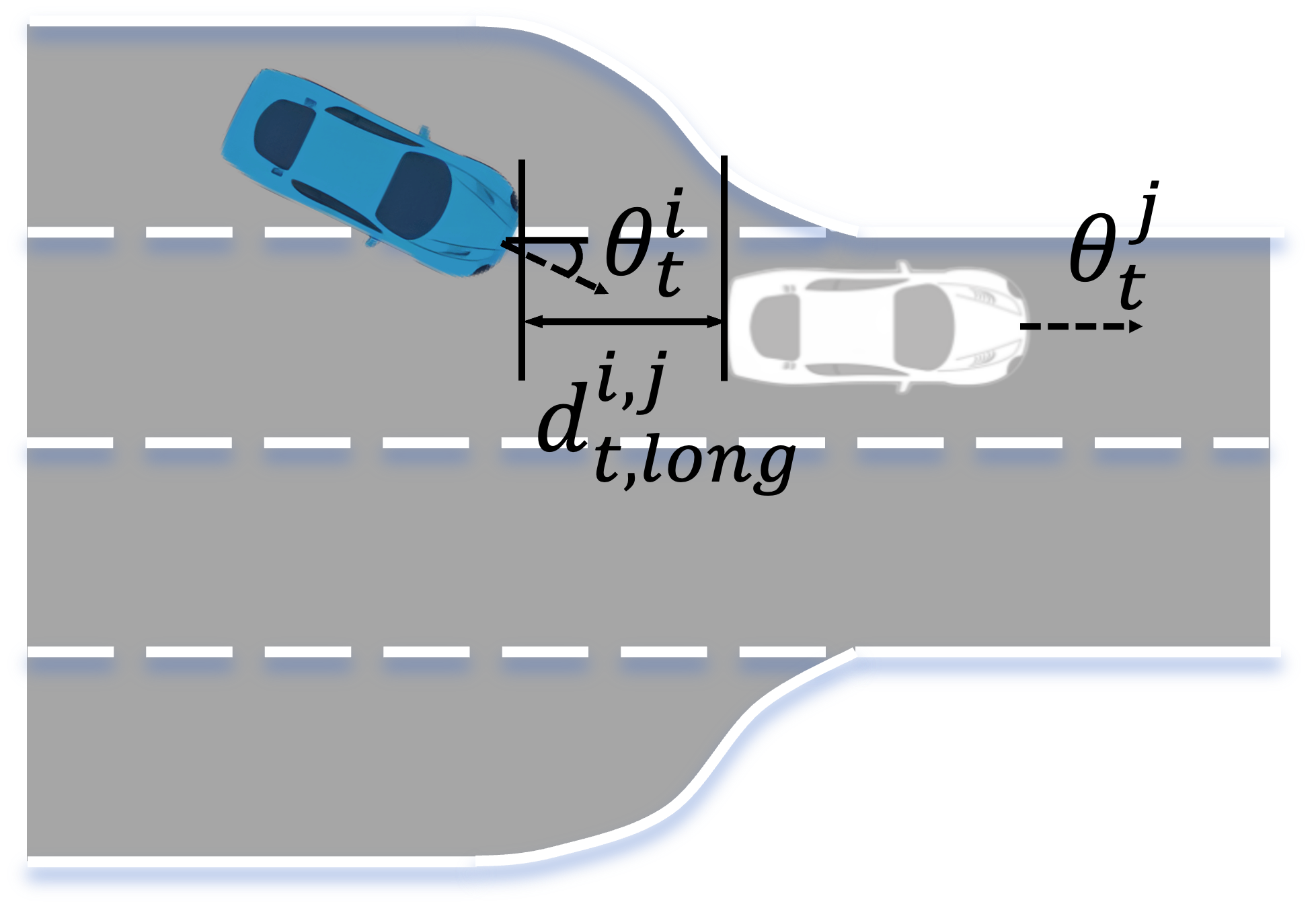}
\caption{Improvement of TTC.}
\label{fig:TTC}
\end{figure}

\color{black}
\begin{algorithm}[htpb]
\caption{{\color{black}Decentralized} Interaction-Adaptive Decision-Making (D-IADM) for CAVs}
\label{alg:D-IADM}
\begin{algorithmic}[1]
\State \textbf{Initialize:} Local observations $o_0^1, \ldots, o_0^n$, $\text{done} = \text{False}$
\For{$t = 1$ to $L$}
        \State Update observations $o_t^1, \ldots, o_t^n$
        \For{$i = 1$ to $n$}
        \State Obtain HDV interaction features $\mathbf{f}_t^{H,i}$ using MH-GAT
        \State Obtain CAV interaction features $\mathbf{f}_t^{C,i}$ using MH-GAT
        \State Obtain context-aware traffic features $\mathbf{f}_t^{dt,i}$ using MLP
        \State Fuse features to get comprehensive observation $\mathbf{f}_t^{i}$ using MLP
        \State Concatenate $\mathbf{f}_t^{i}$ with the last action $a_{t-1}^i$
        \State Generate action $\hat{a}_t^i$ using GRU and decision layer
        \If{\textcolor{black}{$\hat{a}_t^i$ violates TTC/gap safety rules}}
                \State \textcolor{black}{Apply heuristic correction to obtain refined action $a_t^i$} (PSAR)
        \Else
                \State Maintain action $a_t^i = \hat{a}_t^i$
        \EndIf
        \EndFor
        \State Execute actions $a_t^1, \ldots, a_t^n$ and receive rewards $r_t^1, \ldots, r_t^n$
        \If{collision occurs or all vehicles exit the road}
        \State $\text{done} = \text{True}$
        \State \textbf{break}
        \EndIf
\EndFor
\end{algorithmic}
\end{algorithm}

\color{black}
PSAR is a rule-based action refinement layer, which takes the actor-proposed action $\hat{a}_t^i \in \mathcal{A}_i$ as input
and outputs a refined action $a_t^i$.
Relative longitudinal gap $d_{t,long}^{i,j}$ and Time-To-Collision (TTC) $\text{TTC}_t^{i,j}$ to the relevant surrounding vehicles serve as risk indicators.
\begin{align}
d_{t,long}^{i,j} &= (x_t^j - x_t^i - L_{veh}) \cos \theta_t^i - \frac{W_{veh}^i}{2}|\sin \theta_t^i| \nonumber \\
\text{TTC}_{t}^{i,j} &= \frac{d_{t,long}^{i,j}}{\Delta v_{t,long}^{i,j}}, \quad j \in \mathcal{M}_{s}^{i} \nonumber \\
\Delta v_{t,long}^{i,j} &= v_{t}^{i} |\cos \theta_t^i| - v_{t}^{j} |\cos \theta_t^j| \nonumber
\end{align}
where $W_{veh}^i$ denotes the width of CAV $i$, and $\Delta v_{t,long}^{i,j}$ represents the longitudinal velocity difference between CAV $i$ and its surrounding vehicle $j \in \mathcal{M}_{s}^{i}$, 
taking into account the vehicles' heading angles.

PSAR applies four action refinements:
\begin{itemize}
    \item \textbf{Cancel-LC}: Cancels the planned lane-change action;
    \item \textbf{Decel-LC}: Maintains the lane-change intent while applying deceleration;
    \item \textbf{Force-Brake}: Overrides the longitudinal command with deceleration;
    \item \textbf{Suppr-Accel}: Suppresses an acceleration command.
\end{itemize}
The complete triggering conditions and speed-reduction formulations are provided in Algorithm~\ref{alg:PSAR}.
Both the actor-generated action $\hat{a}_t^i$ and the PSAR-refined action $a_t^i$ are incorporated into the next time step's observation as action memory.
By retaining both actions, the actor can perceive how its intended decision was modified by the safety module, tracking the discrepancy between planned and executed behaviors throughout training.

\color{black}
In summary, the process of D-IADM for CAVs in mixed traffic is given in \textbf{\textit{Algorithm \ref{alg:D-IADM}}}.
\subsection{Centralized Interaction-Enhanced Critic (C-IEC)}
\textcolor{black}{The C-IEC module is designed to enhance the critic's global traffic understanding capability.
C-IEC explicitly models how vehicle interactions influence traffic dynamics, enabling more accurate value estimation that provides clearer guidance for policy updates.
The key insight is that by understanding the relationship between interactions and traffic evolution, the critic can better assess whether a local cooperative action contributes to global traffic improvement.
The C-IEC module achieves this through its core component, the Integrated Traffic Dynamics Representation (ITDR) module, which constructs a global vehicle interaction graph and employs cross-attention mechanisms to capture the causal relationship between interactions and traffic dynamics.
The ITDR output is then processed through a GRU to capture temporal dependencies, followed by a fully connected layer that outputs value function predictions to guide actor updates.}

\color{black}
\begin{figure}[htpb]
\centering
\includegraphics[width=0.5\textwidth]{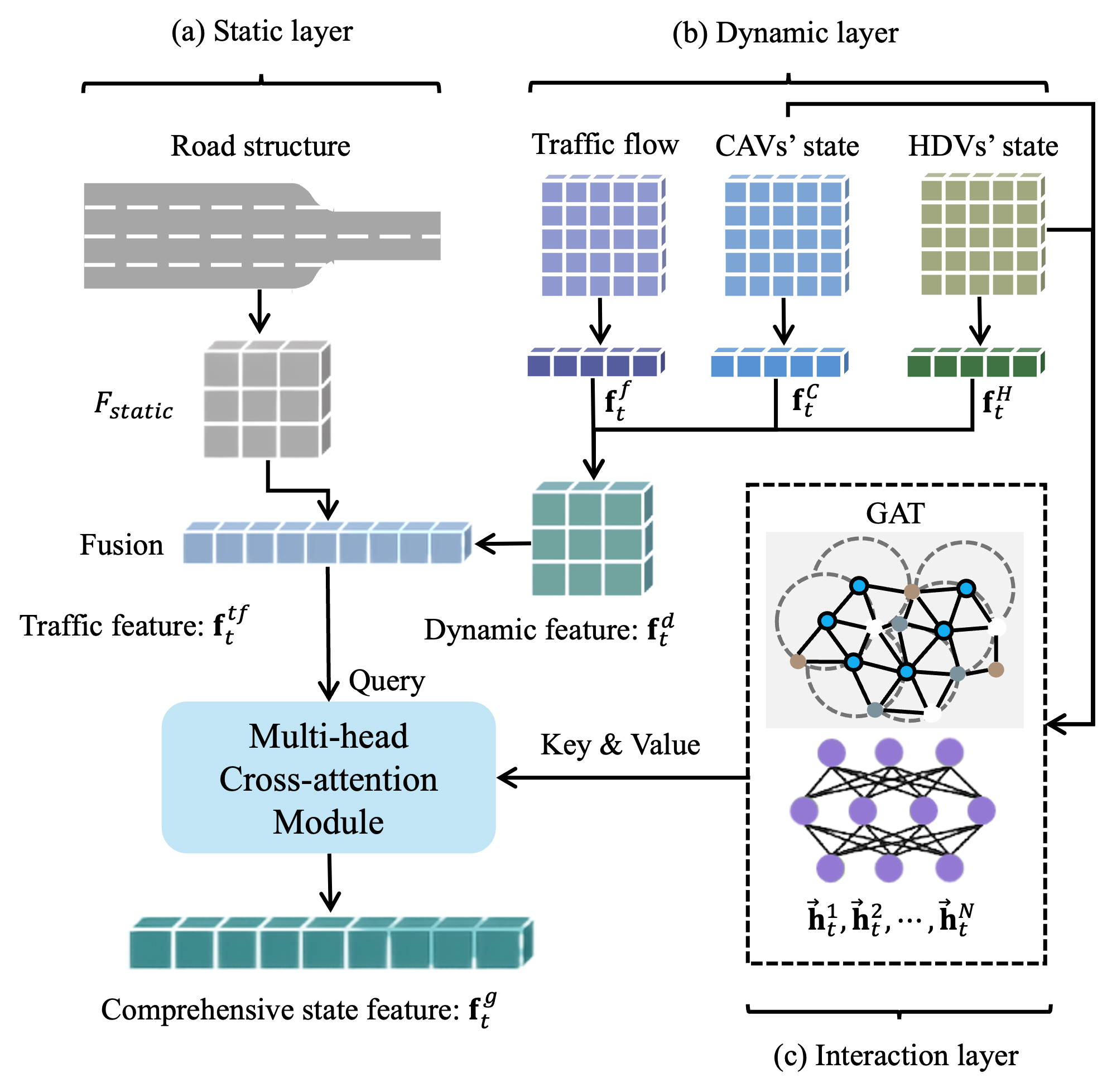}
\caption{Network framework of the ITDR.}
\label{fig:ITDR}
\end{figure}
\textcolor{black}{The ITDR module analyzes the relationship between vehicle interactions and traffic evolution by processing three distinct information layers:
(a) static road structure $\mathbf{F}_{static}$, (b) dynamic traffic information (global lane statistics, CAV and HDV states), and (c) vehicle interaction information.
The first two layers are processed through dedicated MLPs and concatenated to form the traffic feature $\mathbf{f}_{t}^{tf}$.}

\textcolor{black}{For the third layer, vehicle interaction information is processed through a GAT to extract global interaction features.} 
A global vehicle interaction graph, denoted as $\mathcal{G}_{global} = (\mathcal{V}, \mathcal{E})$, is constructed, where $\mathcal{V}$ represents the nodes for all vehicles 
and $\mathcal{E}$ stands for the edges connecting these nodes. 
Here, only adjacent vehicles are connected, i.e., $\mathcal{E} = \{(i, j) | i \in \mathcal{V}, j \in \mathcal{M}_{s}^{i}\}$. 
Each vehicle node's state $m_{t}^{i}=(x_{t}^{i}, y_{t}^{i}, v_{t}^{i}, \theta_{t}^{i})$ is embedded into node features $\mathbf{h}^{i}_{t} \in \mathbb{R}^{S_N}$,
which are fed into the GAT to extract global interaction features $\overrightarrow{\mathbf{h}}_{t}^{i} \in \mathbb{R}^{S_N}$.
The attention weight $\alpha_{ij}$ from node $i$ to node $j$ is calculated as follows:
\begin{equation}
    \alpha_{ij} = \frac{\exp(\text{LeakyReLU}(\overrightarrow{\mathbf{a}} [W\mathbf{h}^i_t \| W\mathbf{h}^j_t]))}{\sum_{k \in \mathcal{M}_{s}^{i}} \exp(\text{LeakyReLU}(\overrightarrow{\mathbf{a}} [W\mathbf{h}^i_t \| W\mathbf{h}^k_t]))}
\end{equation}
where $W$ is the weight matrix, $\overrightarrow{\mathbf{a}}$ is a single-layer feedforward neural network, followed by LeakyReLU nonlinearity and softmax normalization. 
The enhanced node feature $\overrightarrow{\mathbf{h}}_{t}^{i}$ for node $i$ at time $t$ is the weighted sum of all neighboring features:
\begin{equation}
    \overrightarrow{\mathbf{h}}_{t}^{i} = \sigma \left( \sum_{j \in \mathcal{M}_s^i} \alpha_{ij} W \mathbf{h}^j_t \right),
\end{equation}
\textcolor{black}{where $\overrightarrow{\mathbf{h}}_{t} = [\overrightarrow{\mathbf{h}}_{t}^{i}]_{i \in \mathcal{V}}$ denotes the stacked matrix of enhanced node features for all vehicles $i \in \mathcal{V}$.
To capture how vehicle interactions influence traffic evolution, a multi-head cross-attention mechanism is applied with traffic feature $\mathbf{f}_{t}^{tf}$ as Query and interaction feature $\overrightarrow{\mathbf{h}}_{t}$ as Key and Value.
The attention output for each head $h$ is:}
\begin{equation}
\text{Attention}(Q_h, K_h, V_h) = \text{softmax}\left(\frac{Q_h K_h^T}{\sqrt{d_k}}\right)V_h,
\end{equation}
\textcolor{black}{Here $d_k$ is the key dimension, and $Q_h$, $K_h$, $V_h$ are obtained via learned projections:}
\begin{equation}
Q_h = \mathbf{f}_{t}^{tf} W_Q^h, \quad K_h = \overrightarrow{\mathbf{h}}_{t} W_K^h, \quad V_h = \overrightarrow{\mathbf{h}}_{t} W_V^h,
\end{equation}
\textcolor{black}{with $W_Q^h$, $W_K^h$, and $W_V^h$ being the projection matrices for head $h$.
The outputs of all $H$ heads are concatenated and linearly projected:}
\begin{equation}
\text{MultiHead}(Q, K, V) = W_O \left[\text{head}_1; \text{head}_2; \ldots; \text{head}_H \right],
\end{equation}
\textcolor{black}{where $\text{head}_h = \text{Attention}(Q_h, K_h, V_h)$ and $W_O$ is the output projection matrix.}

\textcolor{black}{Through this mechanism, ITDR captures how vehicle interactions influence traffic evolution. Unlike TAIE's local perspective, ITDR provides the critic with a global view of vehicle interactions and traffic dynamics, enabling more accurate value estimation for guiding policy updates.}
\vspace{-10pt}
\section{Experiments}
\color{black}
\subsection{Experimental Setup}

\begin{figure}[htpb]
\centering
\subfloat[\textcolor{black}{25\% reduction}]{\includegraphics[width=\columnwidth]{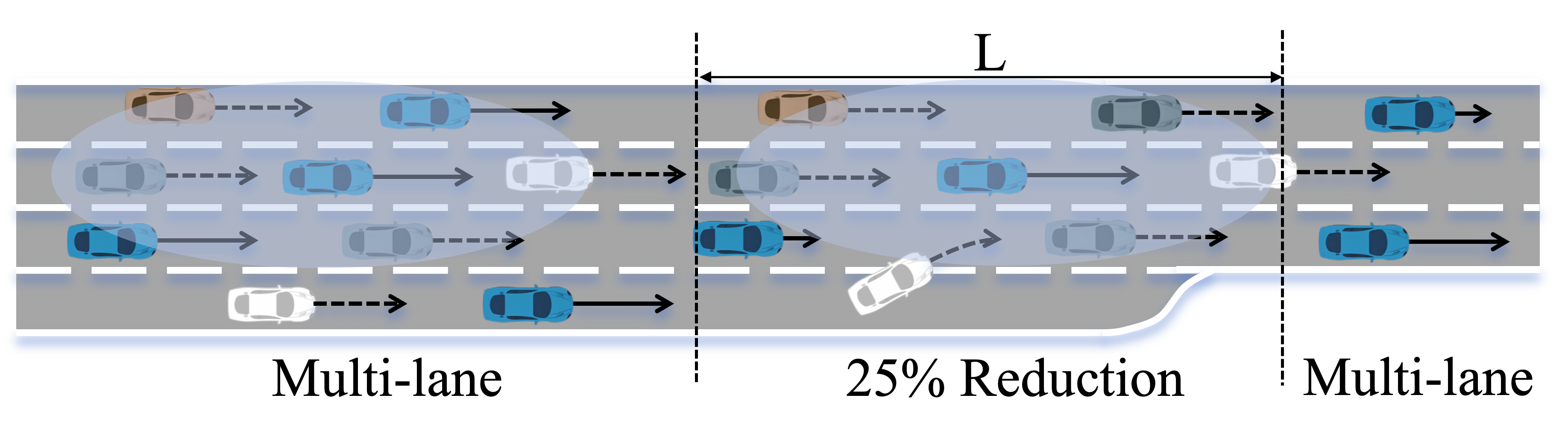}}\\
\subfloat[\textcolor{black}{50\% reduction}]{\includegraphics[width=\columnwidth]{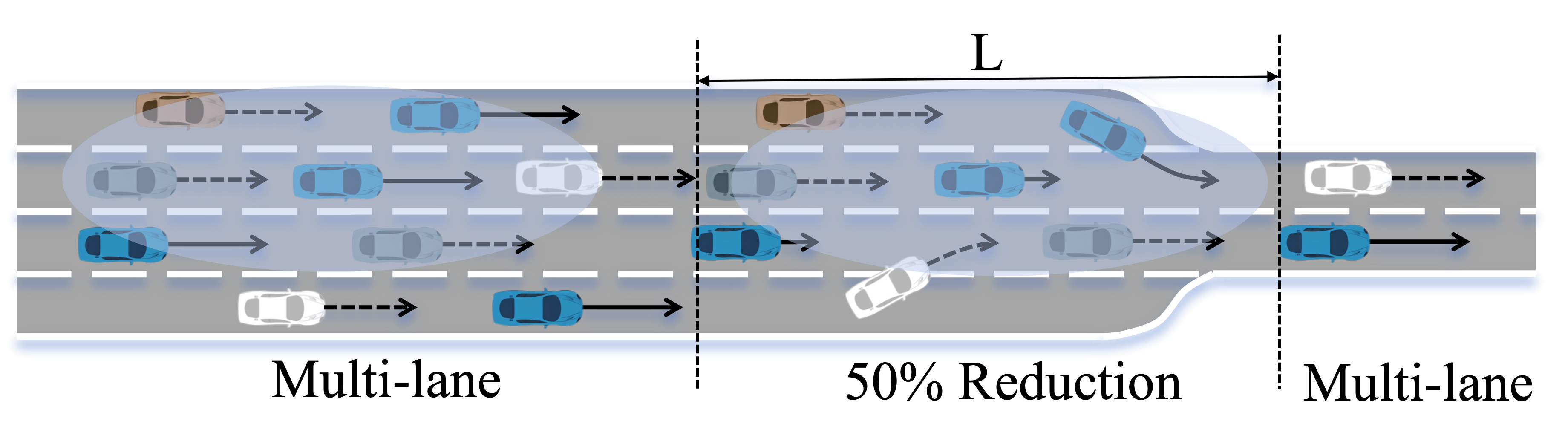}}
\caption{\textcolor{black}{Bottleneck settings with (a) 25\% and (b) 50\% lane reductions}}
\label{fig:scenario_analysis}
\end{figure}

\textcolor{black}{The simulation environment is developed based on SUMO \cite{SUMO} and TransSimHub \cite{TransSimHub}, a unified simulation platform for multi-modal perception and decision-making.
The implementation of this research is developed based on the HARL codebase \cite{kuba2021trust}, and the project repository is available at \texttt{https://github.com/lzxphhh/DIACC.git}.
As illustrated in Fig. \ref{fig:scenario_analysis}, the road network comprises 25\% capacity reduction segments, 50\% capacity reduction segments, and multi-lane segments, forming a 1.3 km long-distance route with a maximum speed limit of 25 m/s.
For generalization testing, an additional 1.3 km test map is designed with reduction segments positioned at different locations from the training map.
HDVs employ the Intelligent Driver Model (IDM) \cite{IDM} for longitudinal control and the LC2013 model \cite{LC2013} for lane-changing maneuvers.}

\textcolor{black}{The experiment employs a systematic training and zero-shot testing framework to evaluate model performance across diverse traffic scenarios.
During training, the scenario contains 25 vehicles with 10 CAVs (penetration rate 0.4) and 15 HDVs following the D1 driving style distribution specified in Table \ref{tab:distribution}.
Three parameter sets represent aggressive, normal, and cautious driving behaviors: aggressive driving features shorter following distances and more frequent lane changes, while cautious driving involves longer following distances and fewer lane changes.
Normal driving adopts default SUMO model parameters.}

\begin{table}[htpb]
\centering
\caption{HDV Driving Style Distributions.}
\begin{tabular}{cccc}
\hline
        Name  & Aggressive & Normal & Cautious \\ \hline
        D1   & 0.2 & 0.6 & 0.2 \\
        D2   & 0.2 & 0.4 & 0.4 \\
        D3   & 0.4 & 0.4 & 0.2 \\ \hline
\end{tabular}
\label{tab:distribution}
\end{table}

\textcolor{black}{For zero-shot generalization testing, the trained model is evaluated on the test map with 
different reduction segment positions and under varied conditions: CAV penetration rates of 0.2 and 0.3, 
vehicle counts of 20, 25, 30 and 40, and HDV driving style distributions D2 and D3.
These tests assess the model's adaptability to unseen road configurations and traffic compositions.}


\subsubsection{Baseline Models}
\textcolor{black}{The comparative experiments encompass MAPPO-based baseline and ablation configurations:}
\begin{itemize}
    \item Pure HDV scenario: All vehicles are HDVs utilizing the IDM+LC2013 model with SUMO's safety mode enabled.
\item SUMO cooperative scenario: CAVs are introduced according to penetration rate, employing the CACC model with highly cooperative lane-changing (lcCooperative=1).
\item MAPPO baseline scenario: CAVs are controlled by the vanilla MAPPO model.
\item \textcolor{black}{Our models: Three MAPPO-based variants are included:}
\begin{itemize}
    \item \textcolor{black}{DIACC (w/o PSAR): DIACC model without the PSAR action refinement module, used for ablation analysis.}
    \item MAPPO-IADM: MAPPO model augmented with only D-IADM module.
    \item DIACC: MAPPO model incorporating both D-IADM and C-IEC modules with PSAR (our primary proposed model).
\end{itemize}
\end{itemize}

In our experiments, we enabled the SUMO simulator's safety mode for HDVs to prevent collisions among HDVs themselves, which would otherwise introduce bias into the evaluation results.

\color{black}
\subsubsection{Training Parameters}
We train the comparative models for 3 million steps in a dense mixed traffic bottleneck scenario with three random seeds. 
The training infrastructure adopts a distributed framework, collecting trajectory data in $20$ parallel environments on CPUs, with a single GPU dedicated to policy learning. 
During training, transition data batches of size $32$ are randomly sampled from the replay buffer and fed into the learner. 
Both actor and critic networks are optimized using the Adam optimizer with the learning rate of $5 \times 10^{-4}$ and discount factor of $0.99$.

\subsection{Results and Analysis}
\color{black}
\subsubsection{Training Performance}

\begin{figure*}[ht]
\centering
\subfloat[Baseline -- Collision Rate]{
        \includegraphics[width=0.31\textwidth]{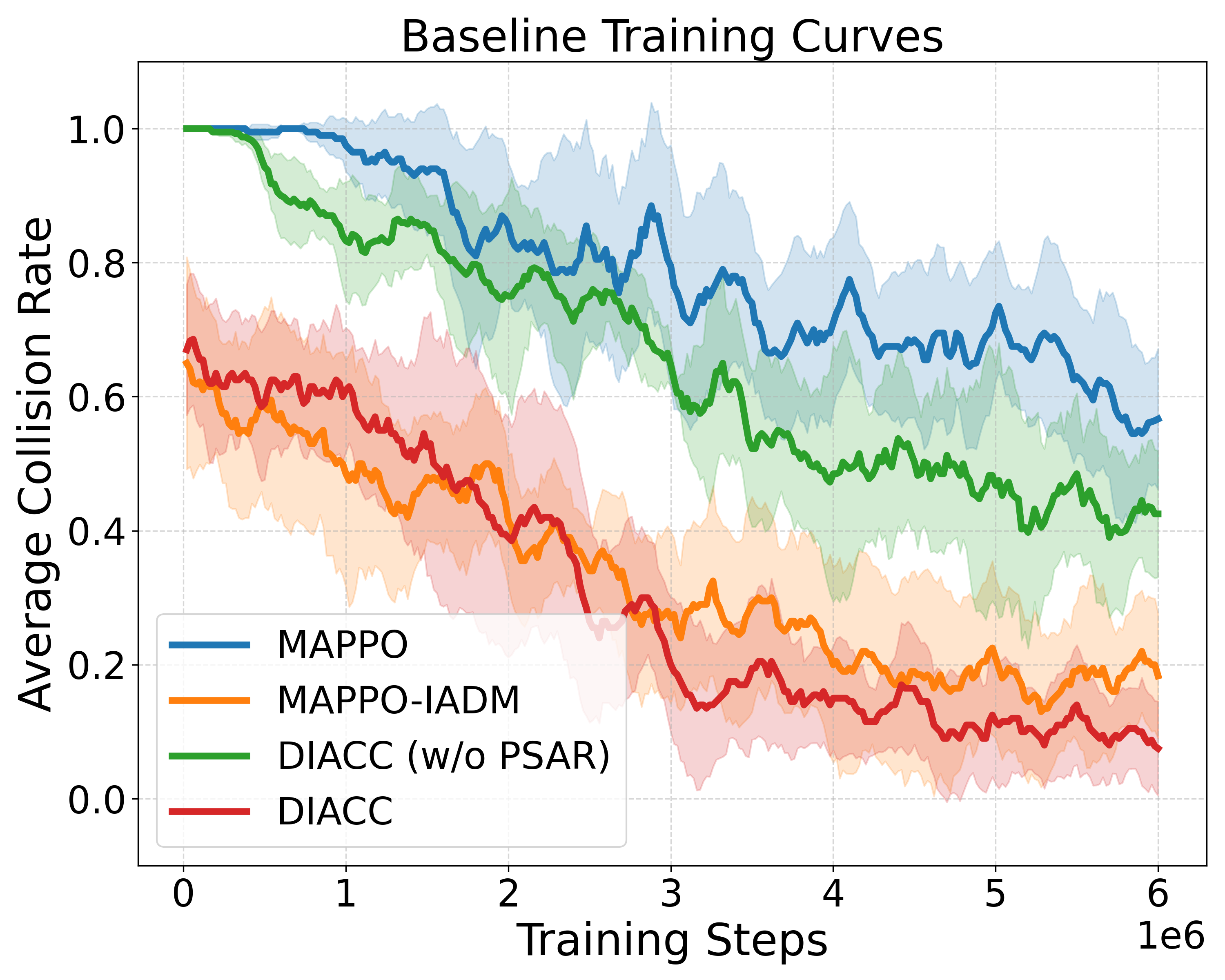}
        \label{fig:train_baseline_collision}
}
\hfill
\subfloat[Baseline -- Global Reward]{
        \includegraphics[width=0.31\textwidth]{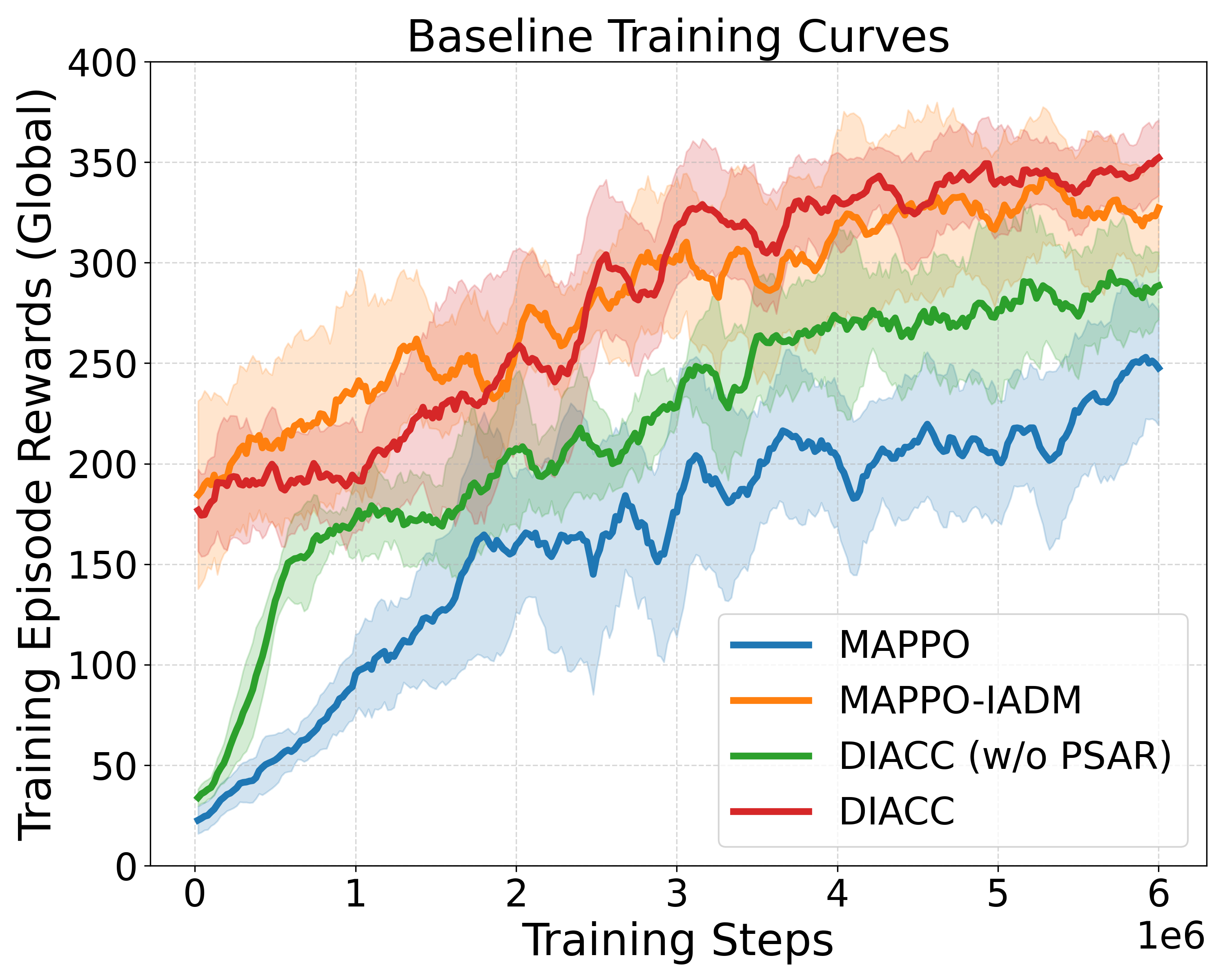}
        \label{fig:train_baseline_global}
}
\hfill
\subfloat[Baseline -- Individual Reward]{
        \includegraphics[width=0.31\textwidth]{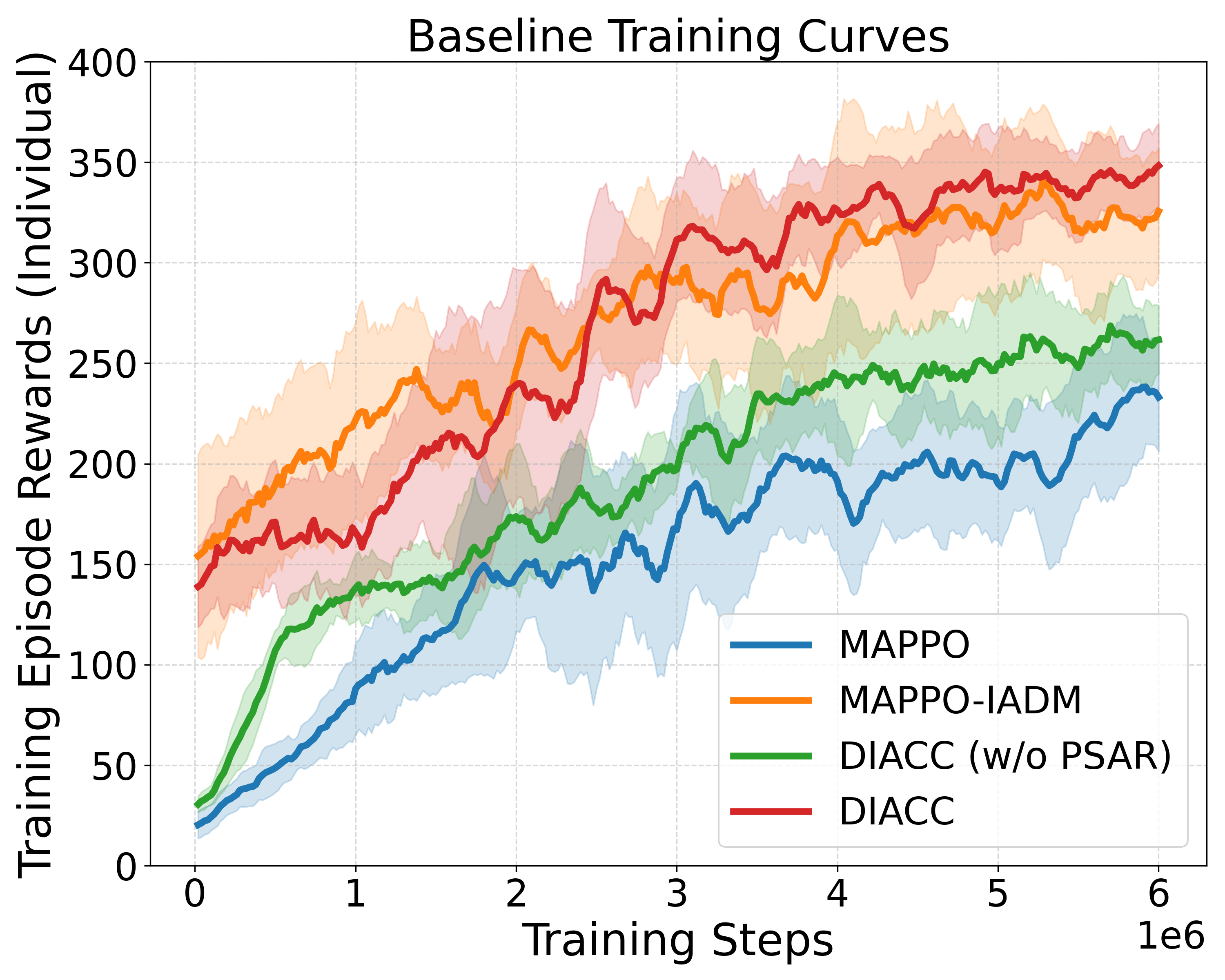}
        \label{fig:train_baseline_individual}
}
\\
\subfloat[$\tau$ selection -- Collision Rate]{
        \includegraphics[width=0.31\textwidth]{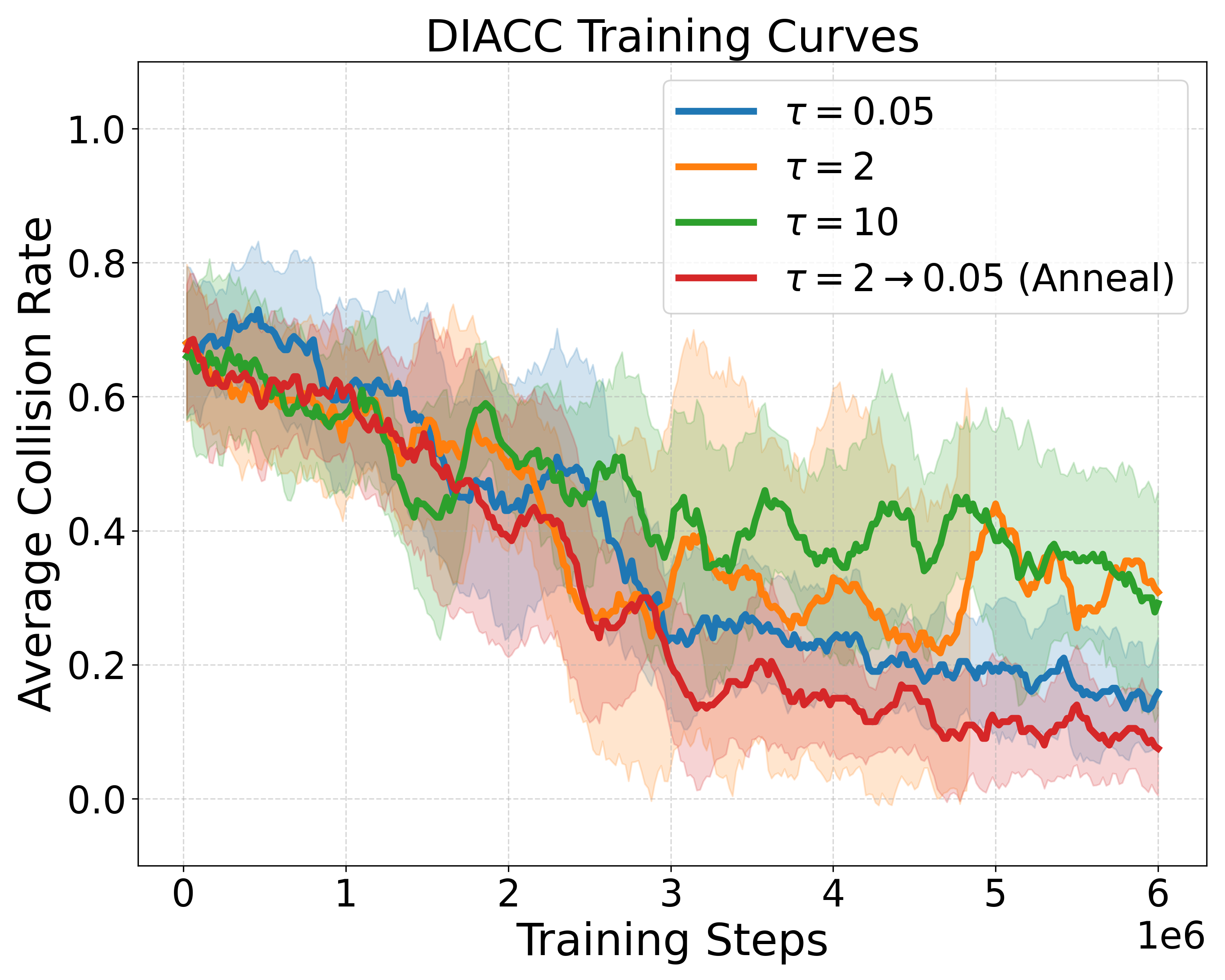}
        \label{fig:train_softmin_collision}
}
\hfill
\subfloat[$\tau$ selection -- Global Reward]{
        \includegraphics[width=0.31\textwidth]{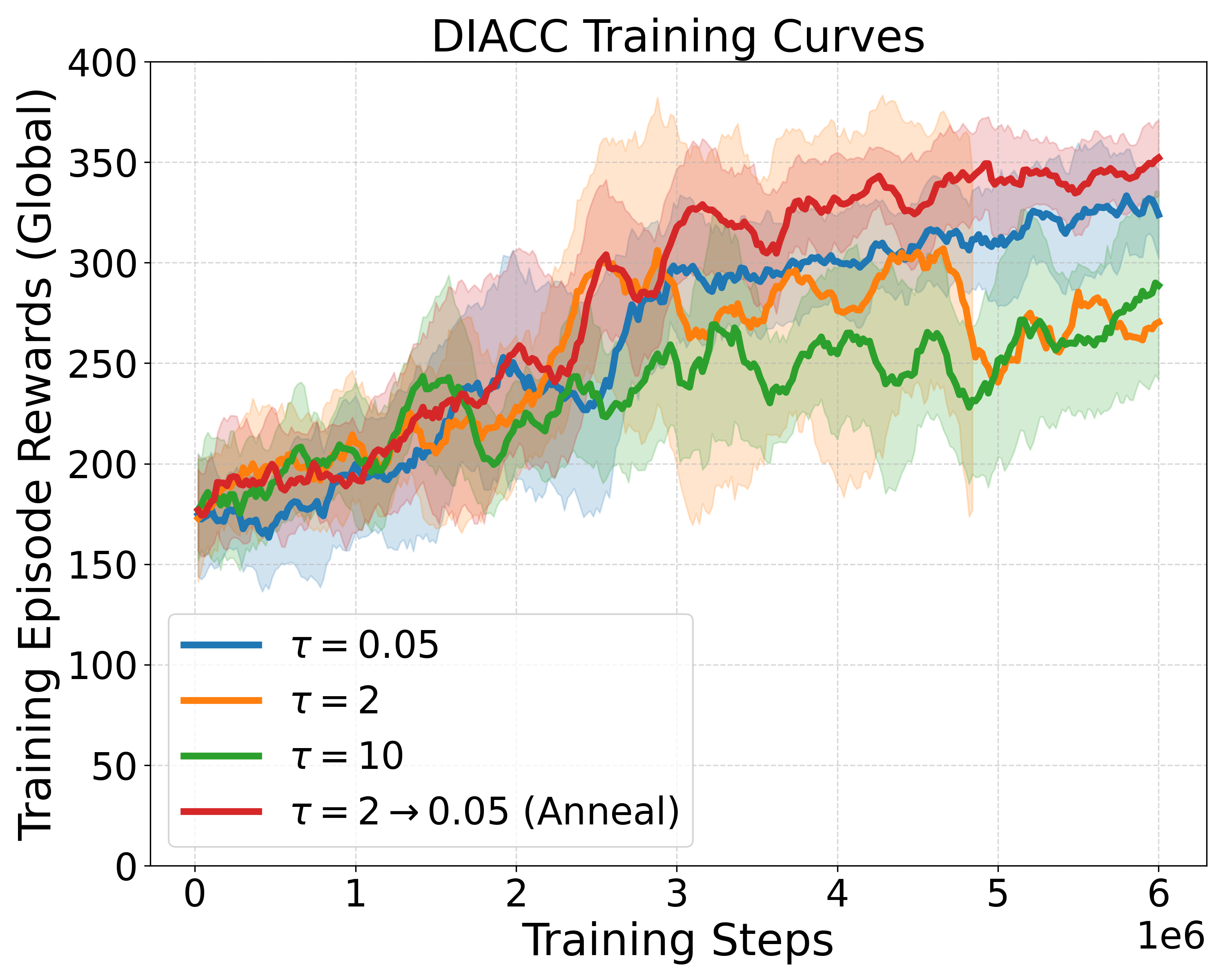}
        \label{fig:train_softmin_global}
}
\hfill
\subfloat[$\tau$ selection -- Individual Reward]{
        \includegraphics[width=0.31\textwidth]{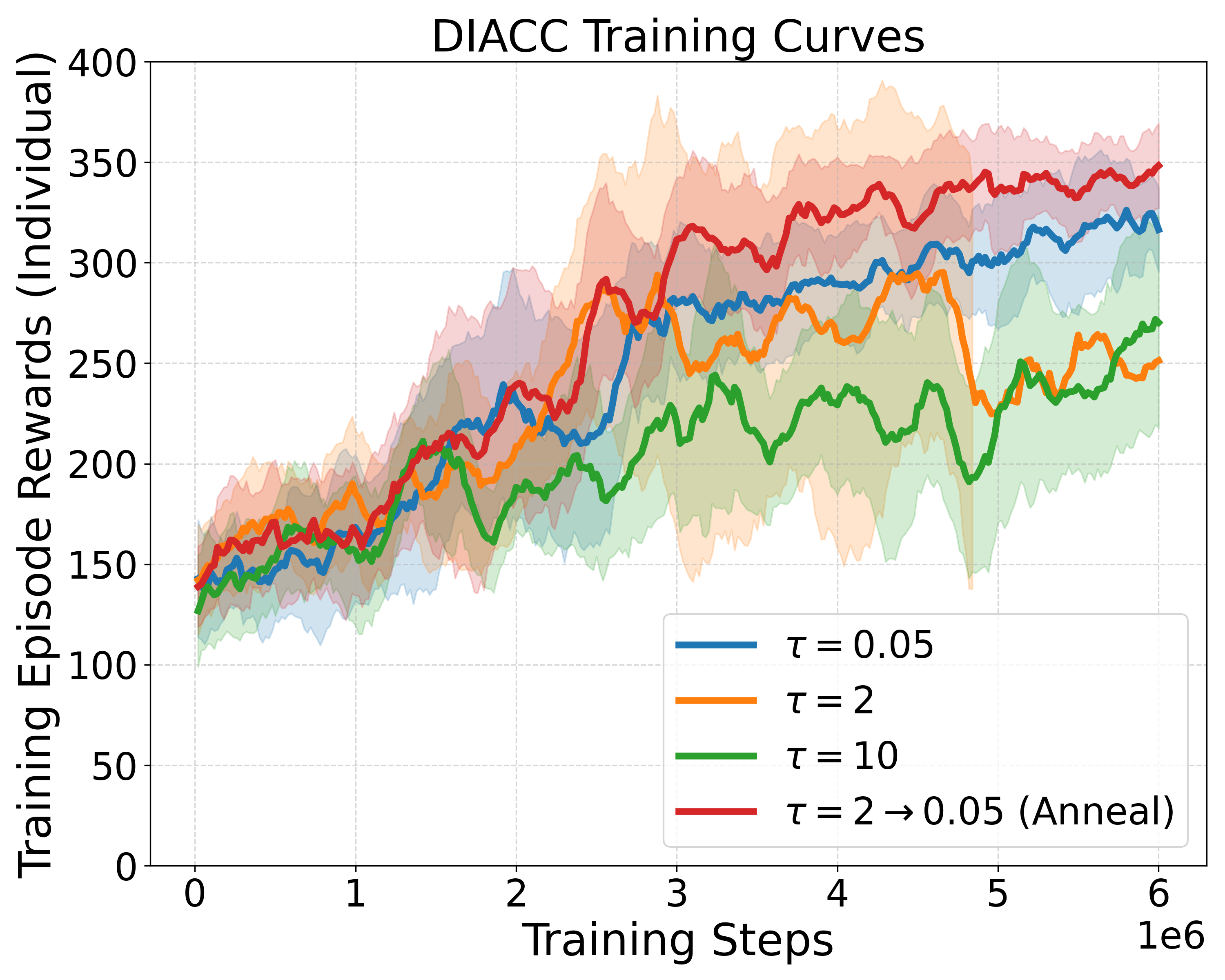}
        \label{fig:train_softmin_individual}
}
\\
\subfloat[$w_e$ selection -- Collision Rate]{
        \includegraphics[width=0.31\textwidth]{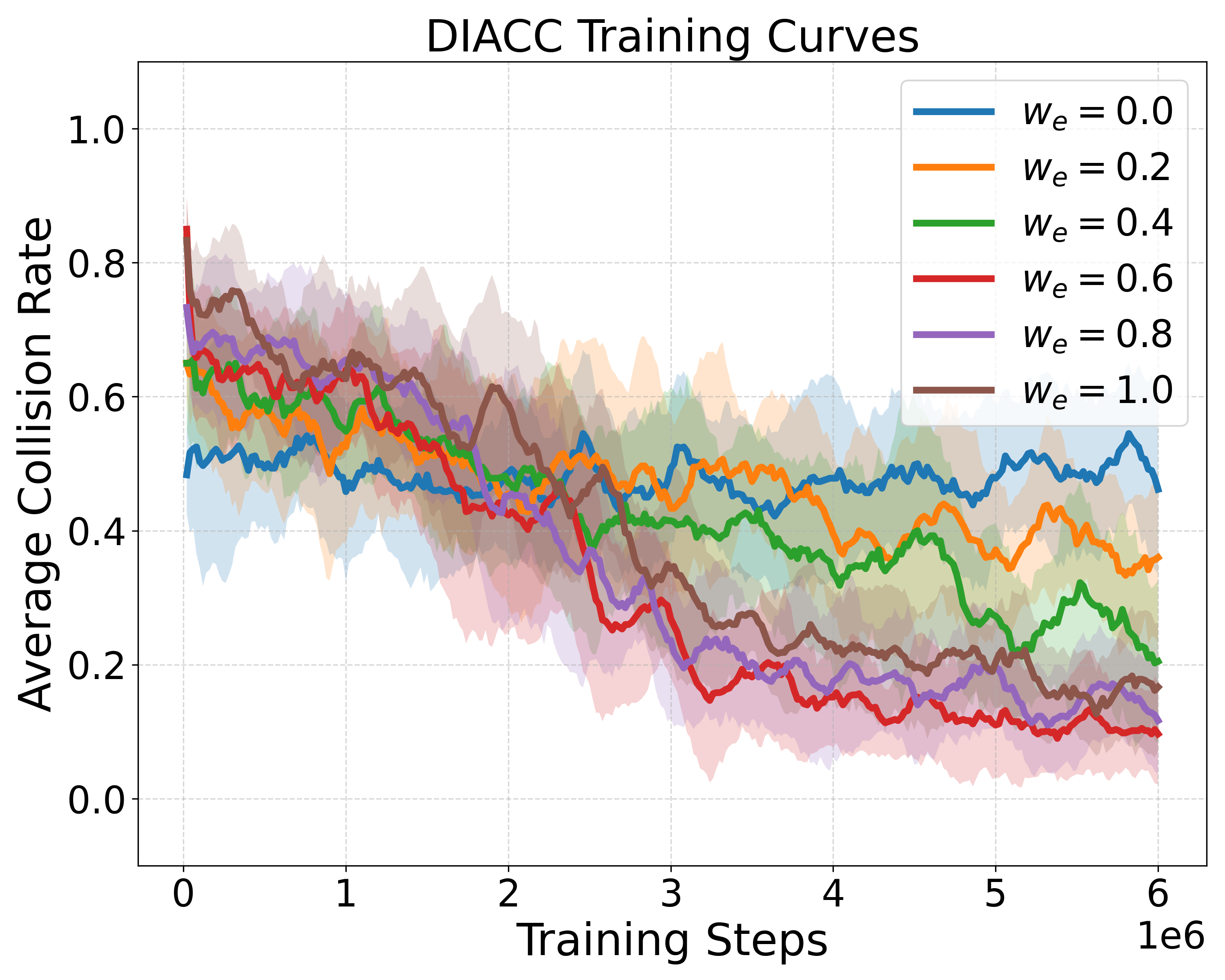}
        \label{fig:train_lambda_collision}
}
\hfill
\subfloat[$w_e$ selection -- Global Reward]{
        \includegraphics[width=0.31\textwidth]{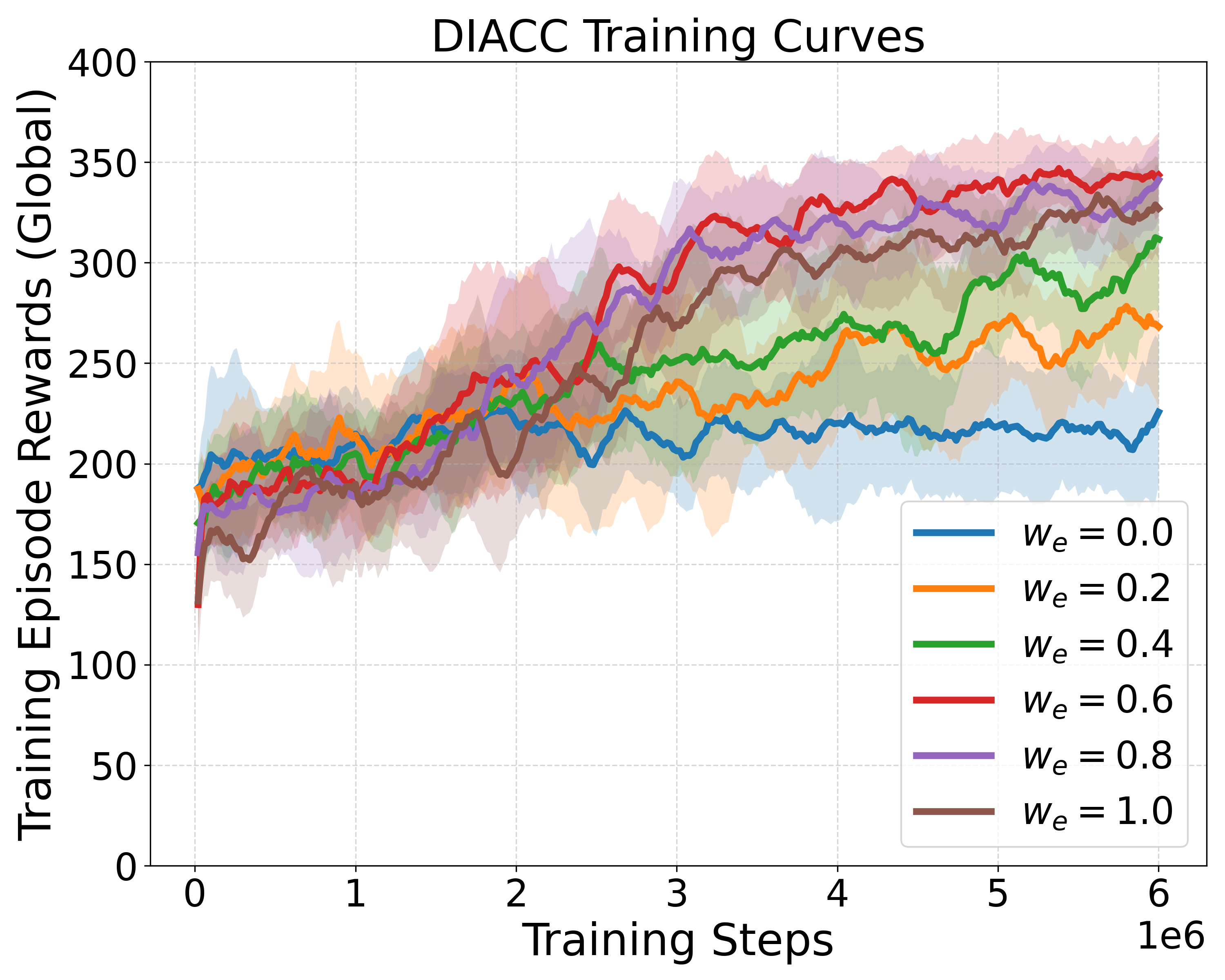}
        \label{fig:train_lambda_global}
}
\hfill
\subfloat[$w_e$ selection -- Individual Reward]{
        \includegraphics[width=0.31\textwidth]{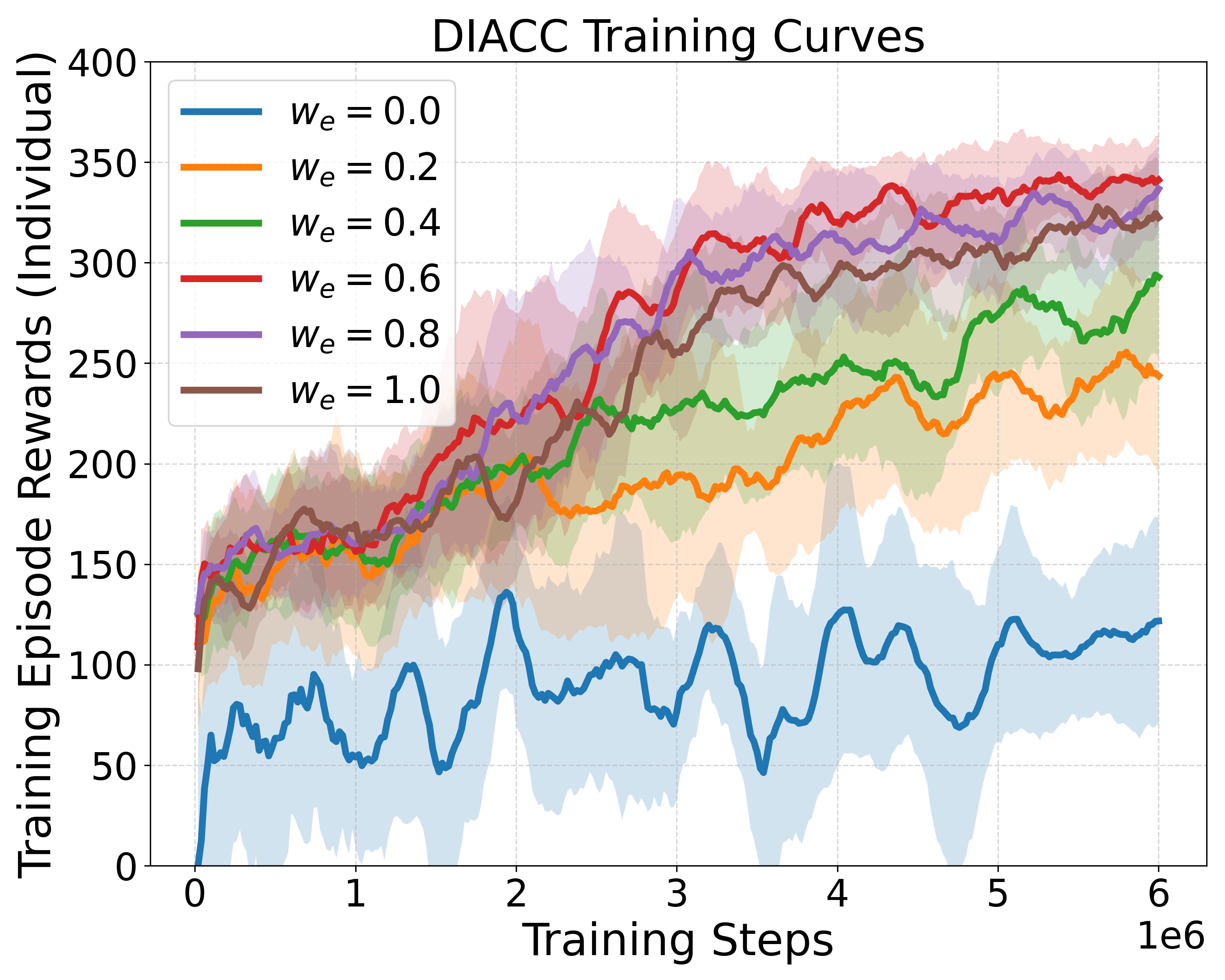}
        \label{fig:train_lambda_individual}
}
\caption{\textcolor{black}{Integrated training-curve comparison. (a)--(c): baseline and ablation comparison. (d)--(f): softmin temperature $\tau$ selection. (g)--(i): ego reward weight $w_e$ selection.}}
\label{fig:training_all}
\end{figure*}

\textcolor{black}{Figs.~\ref{fig:training_all}(a)--(c) compare the training dynamics of baseline and ablation models, including MAPPO, MAPPO-IADM, DIACC (w/o PSAR), and DIACC.
As shown in the figures, DIACC achieves the lowest collision rate and highest reward values at convergence among all compared methods, validating the effectiveness of the full framework.
MAPPO-IADM, which lacks C-IEC, reduces the collision rate more rapidly in early training, yet DIACC surpasses it in later stages.
This behavior confirms that C-IEC, by providing more accurate value estimation through interaction-aware global representations in ITDR, 
guides policy refinement toward superior coordination once basic interaction patterns are established.
DIACC also achieves faster and more stable collision rate reduction than DIACC (w/o PSAR), as PSAR filters high-risk exploratory actions via TTC- and gap-based rules, 
reducing destabilizing unsafe transitions in early training.
Among all variants, vanilla MAPPO performs worst, failing to handle the complex interaction dynamics inherent in mixed traffic.}

\textcolor{black}{Figs.~\ref{fig:training_all}(d)--(f) examine the effect of the temperature parameter $\tau$ by comparing the annealing strategy 
($\tau_{init}=2 \rightarrow \tau_{final}=0.05$ over the first half of training as defined in Eq.~\eqref{eq:tau_annealing}) against fixed-$\tau$ settings.
The annealing strategy achieves the lowest collision rate and highest reward values at convergence, 
demonstrating that the curriculum-inspired schedule effectively balances broad exploration with focused optimization.
Fixed $\tau=0.05$ causes high training variance (wide shaded confidence intervals) and poor collision rate convergence, as the policy is directed toward difficult cases before basic coordination is established.
With fixed $\tau=2$ or $\tau=10$, the softmin weights become increasingly uniform across agents (as $w^i \rightarrow 1/n$ when $\tau \rightarrow \infty$), 
reducing the attention directed toward interaction-intensive scenarios, and training performance degrades monotonically as $\tau$ increases.}

\textcolor{black}{Figs.~\ref{fig:training_all}(g)--(i) examine the effect of the ego reward weight $w_e$ in Eq.~\eqref{eq:reward_all}, which balances local ego objectives and the global coordination reward.
Among all evaluated settings, $w_e=0.6$ achieves the best performance, attaining the lowest collision rate and highest reward values at convergence.
With small $w_e$ values ($w_e=0$ to $0.4$), the ego reward contributes insufficiently to distinguish individual agent performance, 
preventing the policy from selectively directing learning toward interaction-intensive agents, and training performance degrades as $w_e$ decreases.
With large $w_e$ values ($w_e=0.8$ or $w_e=1.0$), both settings yield lower performance than $w_e=0.6$, as the reduced weight on the global reward $r_g$ causes each CAV to over-optimize its local objective,
weakening cooperative coordination and driving the policy toward suboptimal solutions.}

\begin{table*}[!htpb]
\centering
\begin{threeparttable}
\color{black}
\caption{\color{black}Performance Comparison in 25\% Capacity Reduction Bottleneck Scenario.}
\label{tab:table_25_reduction}
\scriptsize
\begin{tabular}{@{}llcccc@{}}
\toprule
\textbf{Scenario} & \textbf{Metric} & \textbf{SUMO (Pure HDV)} & \textbf{SUMO (Mixed Traffic)} & \textbf{MAPPO-IADM} & \textbf{DIACC} \\
\midrule
\multicolumn{6}{c}{\textit{Training Scenario}} \\
\midrule
\multirow{3}{*}{N=25, $\rho=0.4$, D1} & \textit{Speed (m/s)} & $16.08 \pm 7.94$ & $17.05 \pm 7.24$ (+6.0\%) & \textbf{$\boldsymbol{20.77 \pm 2.88}$ (+29.2\%)} & \cellcolor{gray!20}\underline{$19.04 \pm 6.19$ (+18.4\%)} \\
 & \textit{p(WEs)} & $11.20 \pm 4.66\%$ & $8.80 \pm 7.76\%$ (-21.4\%) & \textbf{$\boldsymbol{0.00 \pm 0.00\%}$ (-100.0\%)} & \cellcolor{gray!20}\underline{$4.00 \pm 0.00\%$ (-64.3\%)} \\
 & \textit{p(SCEs)} & $7.20 \pm 3.92\%$ & $6.40 \pm 9.33\%$ (-11.1\%) & \textbf{$\boldsymbol{0.00 \pm 0.00\%}$ (-100.0\%)} & \cellcolor{gray!20}\textbf{$\boldsymbol{0.00 \pm 0.00\%}$ (-100.0\%)} \\
\midrule
\multicolumn{6}{c}{\textit{Zero-shot Testing Scenarios}} \\
\midrule
\multirow{3}{*}{N=25, $\rho=0.4$, D2} & \textit{Speed (m/s)} & $14.44 \pm 8.74$ & $14.54 \pm 9.07$ (+0.7\%) & \textbf{$\boldsymbol{20.13 \pm 4.62}$ (+39.4\%)} & \cellcolor{gray!20}\underline{$19.05 \pm 6.61$ (+31.9\%)} \\
 & \textit{p(WEs)} & $18.40 \pm 4.08\%$ & $11.20 \pm 3.92\%$ (-39.1\%) & \textbf{$\boldsymbol{4.00 \pm 0.00\%}$ (-78.3\%)} & \cellcolor{gray!20}\textbf{$\boldsymbol{4.00 \pm 0.00\%}$ (-78.3\%)} \\
 & \textit{p(SCEs)} & $6.40 \pm 3.20\%$ & $5.60 \pm 6.97\%$ (-12.5\%) & \textbf{$\boldsymbol{0.00 \pm 0.00\%}$ (-100.0\%)} & \cellcolor{gray!20}\textbf{$\boldsymbol{0.00 \pm 0.00\%}$ (-100.0\%)} \\
\cmidrule{1-6}
\multirow{3}{*}{N=25, $\rho=0.4$, D3} & \textit{Speed (m/s)} & $16.48 \pm 7.90$ & $16.12 \pm 8.73$ (-2.2\%) & \underline{$18.86 \pm 6.91$ (+14.4\%)} & \cellcolor{gray!20}\textbf{$\boldsymbol{20.47 \pm 3.64}$ (+24.2\%)} \\
 & \textit{p(WEs)} & $11.20 \pm 5.31\%$ & \underline{$10.61 \pm 5.29\%$ (-5.3\%)} & $12.00 \pm 0.00\%$ (+7.1\%) & \cellcolor{gray!20}\textbf{$\boldsymbol{0.00 \pm 0.00\%}$ (-100.0\%)} \\
 & \textit{p(SCEs)} & \underline{$5.60 \pm 6.97\%$} & $12.42 \pm 12.31\%$ (+121.7\%) & $8.00 \pm 0.00\%$ (+42.9\%) & \cellcolor{gray!20}\textbf{$\boldsymbol{0.00 \pm 0.00\%}$ (-100.0\%)} \\
\cmidrule{1-6}
\multirow{3}{*}{N=20, $\rho=0.4$, D1} & \textit{Speed (m/s)} & $16.26 \pm 8.04$ & $17.54 \pm 8.58$ (+7.9\%) & \textbf{$\boldsymbol{21.36 \pm 2.61}$ (+31.4\%)} & \cellcolor{gray!20}\underline{$19.26 \pm 6.60$ (+18.5\%)} \\
 & \textit{p(WEs)} & $15.00 \pm 5.48\%$ & $9.00 \pm 2.00\%$ (-40.0\%) & \textbf{$\boldsymbol{0.00 \pm 0.00\%}$ (-100.0\%)} & \cellcolor{gray!20}\underline{$5.00 \pm 0.00\%$ (-66.7\%)} \\
 & \textit{p(SCEs)} & $4.00 \pm 8.00\%$ & $3.00 \pm 6.00\%$ (-25.0\%) & \textbf{$\boldsymbol{0.00 \pm 0.00\%}$ (-100.0\%)} & \cellcolor{gray!20}\textbf{$\boldsymbol{0.00 \pm 0.00\%}$ (-100.0\%)} \\
\cmidrule{1-6}
\multirow{3}{*}{N=30, $\rho=0.3$, D1} & \textit{Speed (m/s)} & $15.24 \pm 8.26$ & $16.77 \pm 7.42$ (+10.1\%) & \textbf{$\boldsymbol{18.65 \pm 5.19}$ (+22.4\%)} & \cellcolor{gray!20}\underline{$18.48 \pm 5.20$ (+21.3\%)} \\
 & \textit{p(WEs)} & $12.00 \pm 4.99\%$ & $7.33 \pm 3.89\%$ (-38.9\%) & \textbf{$\boldsymbol{3.33 \pm 0.00\%}$ (-72.2\%)} & \cellcolor{gray!20}\textbf{$\boldsymbol{3.33 \pm 0.00\%}$ (-72.2\%)} \\
 & \textit{p(SCEs)} & $8.00 \pm 7.77\%$ & $2.00 \pm 4.00\%$ (-75.0\%) & \underline{$1.33 \pm 2.67\%$ (-83.3\%)} & \cellcolor{gray!20}\textbf{$\boldsymbol{0.00 \pm 0.00\%}$ (-100.0\%)} \\
\cmidrule{1-6}
\multirow{3}{*}{N=40, $\rho=0.2$, D1} & \textit{Speed (m/s)} & $15.03 \pm 8.11$ & $14.52 \pm 8.81$ (-3.4\%) & \underline{$15.63 \pm 7.16$ (+4.0\%)} & \cellcolor{gray!20}\textbf{$\boldsymbol{17.24 \pm 7.70}$ (+14.7\%)} \\
 & \textit{p(WEs)} & $10.00 \pm 3.16\%$ & $13.00 \pm 4.58\%$ (+30.0\%) & \textbf{$\boldsymbol{5.00 \pm 0.00\%}$ (-50.0\%)} & \cellcolor{gray!20}\underline{$7.50 \pm 0.00\%$ (-25.0\%)} \\
 & \textit{p(SCEs)} & $8.00 \pm 2.92\%$ & $15.00 \pm 8.51\%$ (+87.5\%) & \textbf{$\boldsymbol{0.00 \pm 0.00\%}$ (-100.0\%)} & \cellcolor{gray!20}\textbf{$\boldsymbol{0.00 \pm 0.00\%}$ (-100.0\%)} \\
\bottomrule
\end{tabular}
\end{threeparttable}
\end{table*}

\begin{table*}[!htpb]
\centering
\begin{threeparttable}
\color{black}
\caption{\color{black}Performance Comparison in 50\% Capacity Reduction Bottleneck Scenario.}
\label{tab:table_50_reduction}
\scriptsize
\begin{tabular}{@{}llcccc@{}}
\toprule
\textbf{Scenario} & \textbf{Metric} & \textbf{SUMO (Pure HDV)} & \textbf{SUMO (Mixed Traffic)} & \textbf{MAPPO-IADM} & \textbf{DIACC} \\
\midrule
\multicolumn{6}{c}{\textit{Training Scenario}} \\
\midrule
\multirow{3}{*}{N=25, $\rho=0.4$, D1} & \textit{Speed (m/s)} & $10.00 \pm 7.30$ & $14.10 \pm 6.52$ (+41.0\%) & \underline{$17.47 \pm 6.65$ (+74.8\%)} & \cellcolor{gray!20}\textbf{$\boldsymbol{17.60 \pm 7.77}$ (+76.0\%)} \\
 & \textit{p(WEs)} & $25.60 \pm 9.83\%$ & $13.23 \pm 7.97\%$ (-48.3\%) & \textbf{$\boldsymbol{6.00 \pm 2.00\%}$ (-76.6\%)} & \cellcolor{gray!20}\underline{$6.80 \pm 5.38\%$ (-73.4\%)} \\
 & \textit{p(SCEs)} & $16.40 \pm 7.68\%$ & $7.60 \pm 8.48\%$ (-53.7\%) & \underline{$6.40 \pm 7.84\%$ (-61.0\%)} & \cellcolor{gray!20}\textbf{$\boldsymbol{0.00 \pm 0.00\%}$ (-100.0\%)} \\
\midrule
\multicolumn{6}{c}{\textit{Zero-shot Testing Scenarios}} \\
\midrule
\multirow{3}{*}{N=25, $\rho=0.4$, D2} & \textit{Speed (m/s)} & $10.55 \pm 7.71$ & $13.21 \pm 6.67$ (+25.2\%) & \underline{$16.47 \pm 7.96$ (+56.1\%)} & \cellcolor{gray!20}\textbf{$\boldsymbol{16.59 \pm 8.48}$ (+57.2\%)} \\
 & \textit{p(WEs)} & $23.60 \pm 9.37\%$ & \underline{$10.90 \pm 6.96\%$ (-53.8\%)} & $12.00 \pm 0.00\%$ (-49.2\%) & \cellcolor{gray!20}\textbf{$\boldsymbol{10.00 \pm 6.00\%}$ (-57.6\%)} \\
 & \textit{p(SCEs)} & $8.80 \pm 6.65\%$ & \textbf{$\boldsymbol{3.67 \pm 5.26\%}$ (-58.3\%)} & $8.00 \pm 0.00\%$ (-9.1\%) & \cellcolor{gray!20}\underline{$4.00 \pm 4.00\%$} (-54.5\%) \\
\cmidrule{1-6}
\multirow{3}{*}{N=25, $\rho=0.4$, D3} & \textit{Speed (m/s)} & $12.65 \pm 7.80$ & $14.98 \pm 7.47$ (+18.5\%) & \underline{$15.32 \pm 8.05$ (+21.1\%)} & \cellcolor{gray!20}\textbf{$\boldsymbol{15.90 \pm 7.78}$ (+25.7\%)} \\
 & \textit{p(WEs)} & $16.40 \pm 5.20\%$ & \underline{$10.03 \pm 8.41\%$ (-38.8\%)} & $12.00 \pm 4.00\%$ (-26.8\%) & \cellcolor{gray!20}\textbf{$\boldsymbol{10.00 \pm 2.00\%}$ (-39.0\%)} \\
 & \textit{p(SCEs)} & $9.20 \pm 7.60\%$ & \textbf{$\boldsymbol{3.20 \pm 6.40\%}$ (-65.2\%)} & $14.00 \pm 2.00\%$ (+52.2\%) & \cellcolor{gray!20}\underline{$8.00 \pm 0.00\%$} (-13.0\%) \\
\cmidrule{1-6}
\multirow{3}{*}{N=20, $\rho=0.4$, D1} & \textit{Speed (m/s)} & $11.99 \pm 7.64$ & $14.91 \pm 6.30$ (+24.3\%) & \textbf{$\boldsymbol{17.98 \pm 5.78}$ (+50.0\%)} & \cellcolor{gray!20}\underline{$17.62 \pm 7.61$ (+46.9\%)} \\
 & \textit{p(WEs)} & $19.00 \pm 8.31\%$ & $10.00 \pm 4.47\%$ (-47.4\%) & \textbf{$\boldsymbol{5.00 \pm 0.00\%}$ (-73.7\%)} & \cellcolor{gray!20}\underline{$7.50 \pm 2.50\%$ (-60.5\%)} \\
 & \textit{p(SCEs)} & $13.00 \pm 8.12\%$ & $7.50 \pm 6.80\%$ (-42.3\%) & \textbf{$\boldsymbol{0.00 \pm 0.00\%}$ (-100.0\%)} & \cellcolor{gray!20}\textbf{$\boldsymbol{0.00 \pm 0.00\%}$ (-100.0\%)} \\
\cmidrule{1-6}
\multirow{3}{*}{N=30, $\rho=0.3$, D1} & \textit{Speed (m/s)} & $9.25 \pm 7.36$ & $10.51 \pm 6.79$ (+13.6\%) & \underline{$11.97 \pm 8.10$ (+29.4\%)} & \cellcolor{gray!20}\textbf{$\boldsymbol{14.37 \pm 8.61}$ (+55.3\%)} \\
 & \textit{p(WEs)} & $27.67 \pm 10.96\%$ & $23.67 \pm 7.67\%$ (-14.5\%) & \underline{$16.00 \pm 3.27\%$ (-42.2\%)} & \cellcolor{gray!20}\textbf{$\boldsymbol{13.33 \pm 3.33\%}$ (-51.8\%)} \\
 & \textit{p(SCEs)} & \textbf{$\boldsymbol{20.00 \pm 6.50\%}$} & $20.33 \pm 8.62\%$ (+1.7\%) & $29.00 \pm 7.75\%$ (+45.0\%) & \cellcolor{gray!20}\textbf{\underline{$20.00 \pm 10.00\% (+0.0\%)$}} \\
\cmidrule{1-6}
\multirow{3}{*}{N=40, $\rho=0.2$, D1} & \textit{Speed (m/s)} & $8.40 \pm 6.93$ & $10.49 \pm 7.47$ (+24.8\%) & \underline{$10.84 \pm 7.97$ (+29.0\%)} & \cellcolor{gray!20}\textbf{$\boldsymbol{12.68 \pm 7.93}$ (+51.0\%)} \\
 & \textit{p(WEs)} & $30.75 \pm 9.94\%$ & $24.25 \pm 12.70\%$ (-21.1\%) & \underline{$22.50 \pm 10.00\%$ (-26.8\%)} & \cellcolor{gray!20}\textbf{$\boldsymbol{12.25 \pm 3.61\%}$ (-60.2\%)} \\
 & \textit{p(SCEs)} & $19.00 \pm 7.60\%$ & \underline{$16.00 \pm 12.31\%$ (-15.8\%)} & $23.75 \pm 11.25\%$ (+25.0\%) & \cellcolor{gray!20}\textbf{$\boldsymbol{8.75 \pm 4.37\%}$ (-53.9\%)} \\
\bottomrule
\end{tabular}
\end{threeparttable}
\end{table*}

\vspace{-15pt}

\textcolor{black}{\subsubsection{Testing Performance}
Tables \ref{tab:table_25_reduction} and \ref{tab:table_50_reduction} present comprehensive performance evaluation across 25\% and 50\% capacity reduction bottleneck scenarios, 
where $N$ denotes the total number of vehicles traversing the bottleneck.
Three metrics are evaluated: (1) Speed (m/s), measuring average speed and standard deviation to reflect traffic efficiency, 
(2) p(WEs), the probability of Waiting Events where vehicle speed falls below 3 m/s, and 
(3) p(SCEs), the probability of Safety-Critical Events including proximity violations (distance below minimum safety threshold), TTC below critical limit, collisions, 
and emergency braking \cite{zhou2022developing}.}

\textcolor{black}{Across both Table~\ref{tab:table_25_reduction} and Table~\ref{tab:table_50_reduction}, DIACC consistently achieves optimal or near-optimal results in both efficiency (high speed and low p(WEs)) and safety (low p(SCEs)), 
confirming the effectiveness of the full architecture.
Notably, DIACC reduces p(SCEs) to 0\% across all five zero-shot scenarios in Table~\ref{tab:table_25_reduction}, demonstrating that the complete framework effectively eliminates safety-critical events under diverse driving style distributions.
MAPPO-IADM, which shares the same actor architecture but omits C-IEC, achieves competitive efficiency performance in Table~\ref{tab:table_25_reduction}, even surpassing DIACC in certain scenarios (e.g., N=25, D1: 20.77 vs.\ 19.04~m/s, N=20, D1: 21.36 vs.\ 19.26~m/s).
This indicates that the D-IADM actor design alone suffices for maintaining high efficiency in simple settings.
However, MAPPO-IADM does not consistently achieve zero p(SCEs) in Table~\ref{tab:table_25_reduction} and degrades substantially in the more demanding 50\% reduction scenario, 
particularly in the N=30 and N=40 settings of Table~\ref{tab:table_50_reduction},
where DIACC demonstrates clear advantages in both efficiency and safety.
These contrasting results corroborate the role of C-IEC, which supplies interaction-aware global value estimation during actor training 
to enable the learning of cooperative policies that the actor module alone cannot achieve in high-density, capacity-constrained scenarios.}

\begin{figure*}[ht]
\centering
\subfloat[$N=30$: Speed Distribution]{
        \includegraphics[width=0.48\textwidth]{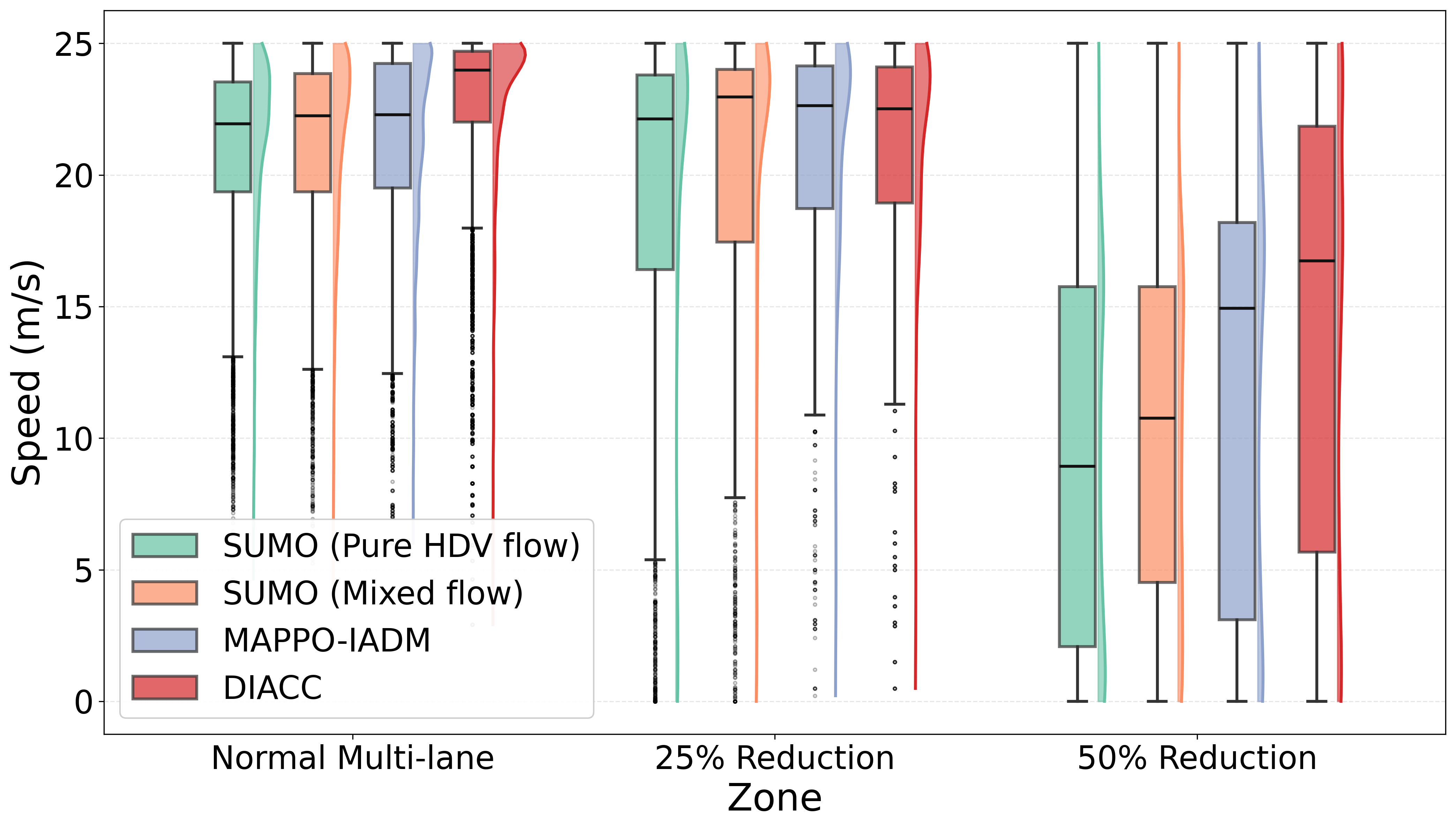}
        \label{fig:speed_boxplot_N30}
}
\subfloat[$N=40$: Speed distributions]{
        \includegraphics[width=0.48\textwidth]{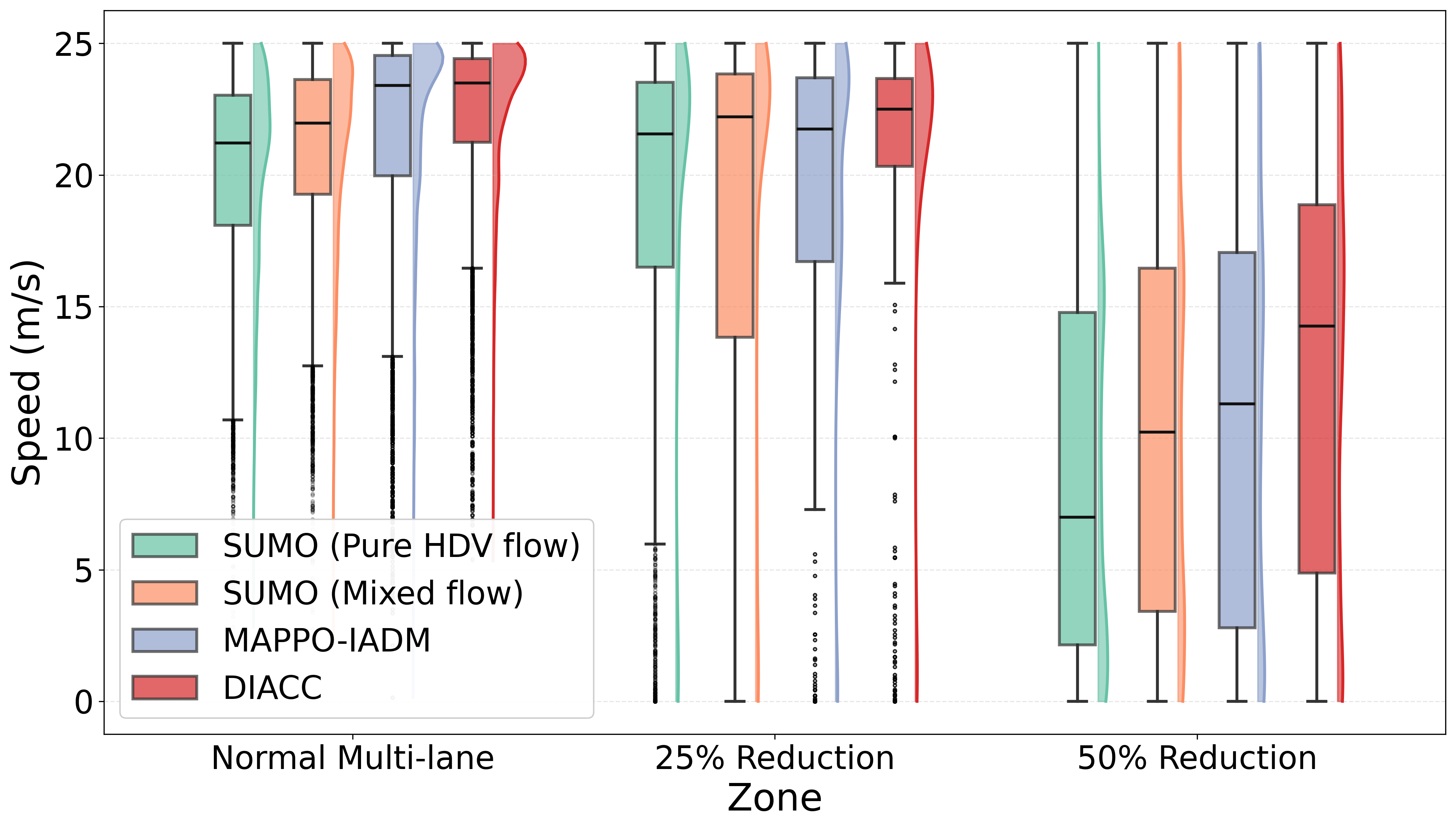}
        \label{fig:speed_boxplot_N40}
}
\\
\subfloat[$N=30$, Normal Multilane Segment]{
        \includegraphics[width=0.31\textwidth]{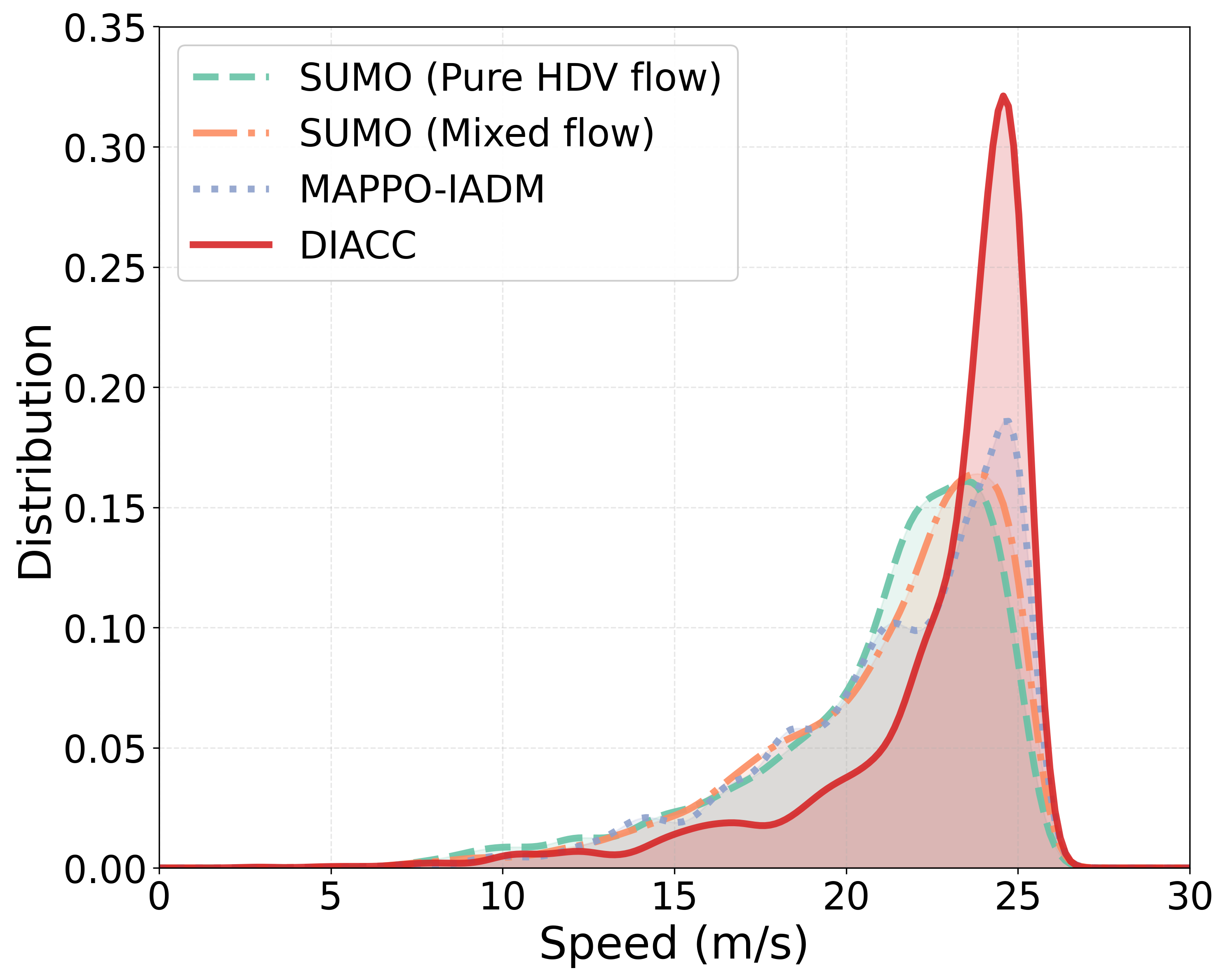}
        \label{fig:speed_kde_N30_normalmultilane}
}
\hfill
\subfloat[$N=30$, 25\% Capacity Reduction]{
        \includegraphics[width=0.31\textwidth]{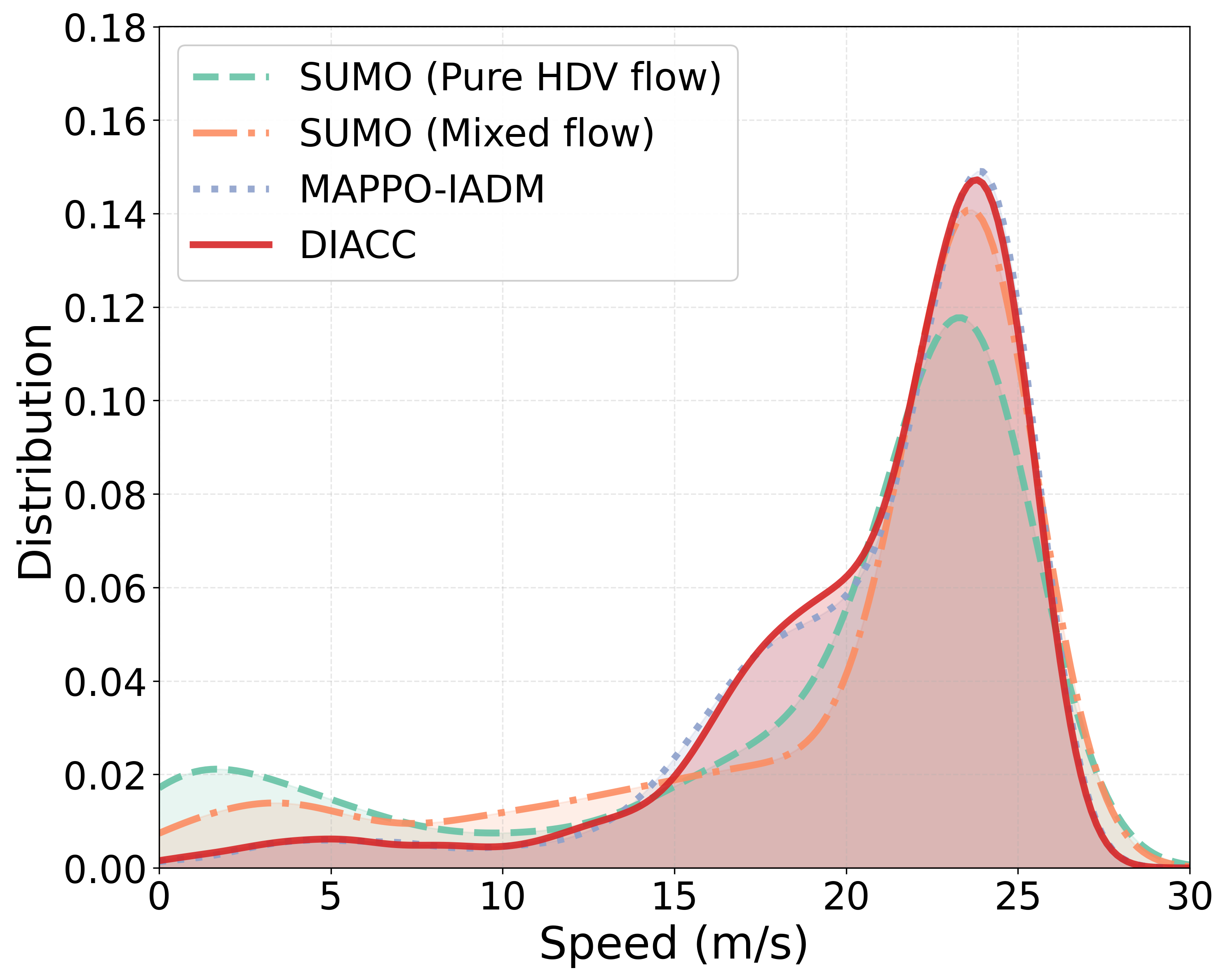}
        \label{fig:speed_kde_N30_25percentreduction}
}
\hfill
\subfloat[$N=30$, 50\% Capacity Reduction]{
        \includegraphics[width=0.31\textwidth]{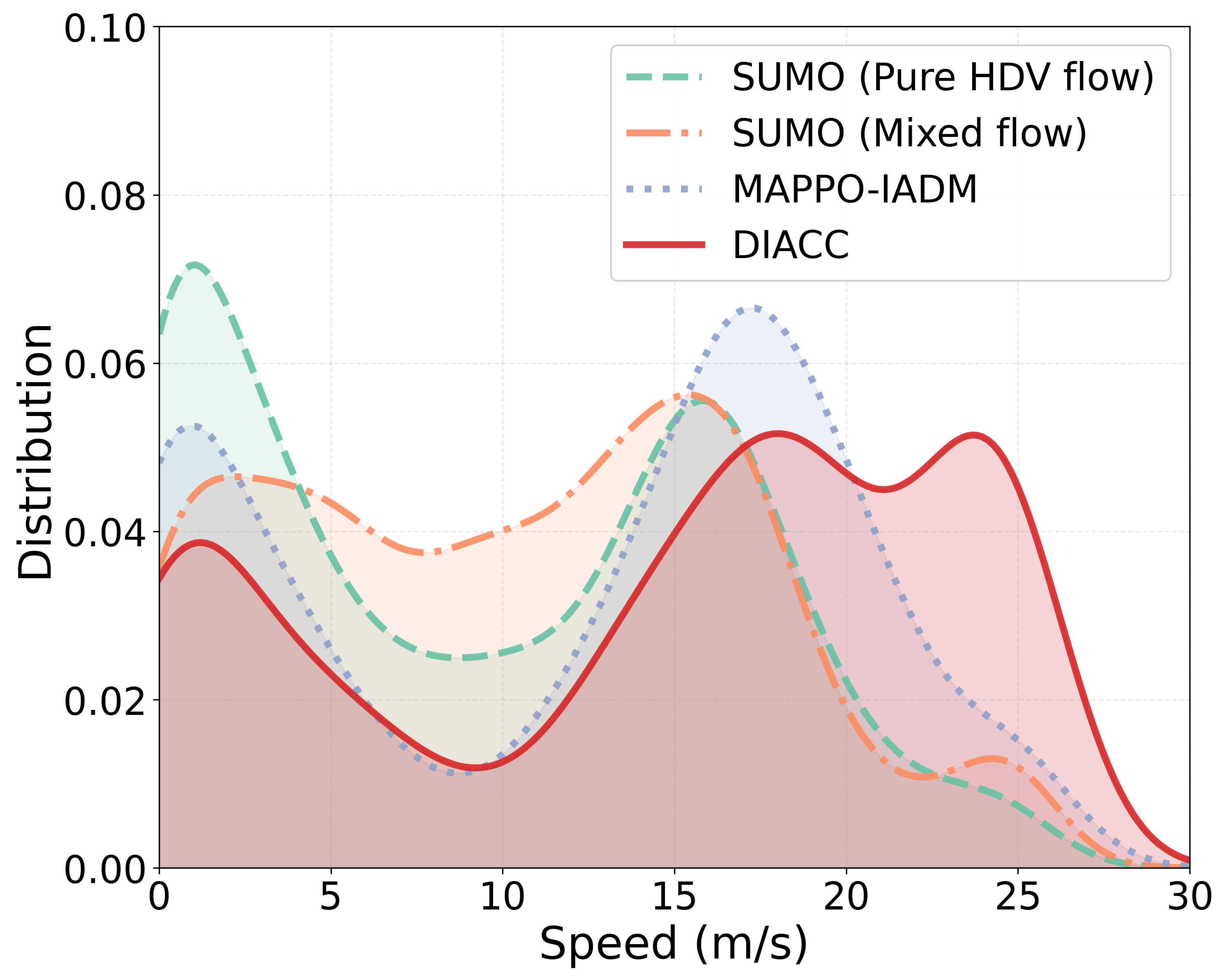}
        \label{fig:speed_kde_N30_50percentreduction}
}
\\
\subfloat[$N=40$, Normal Multilane Segment]{
        \includegraphics[width=0.31\textwidth]{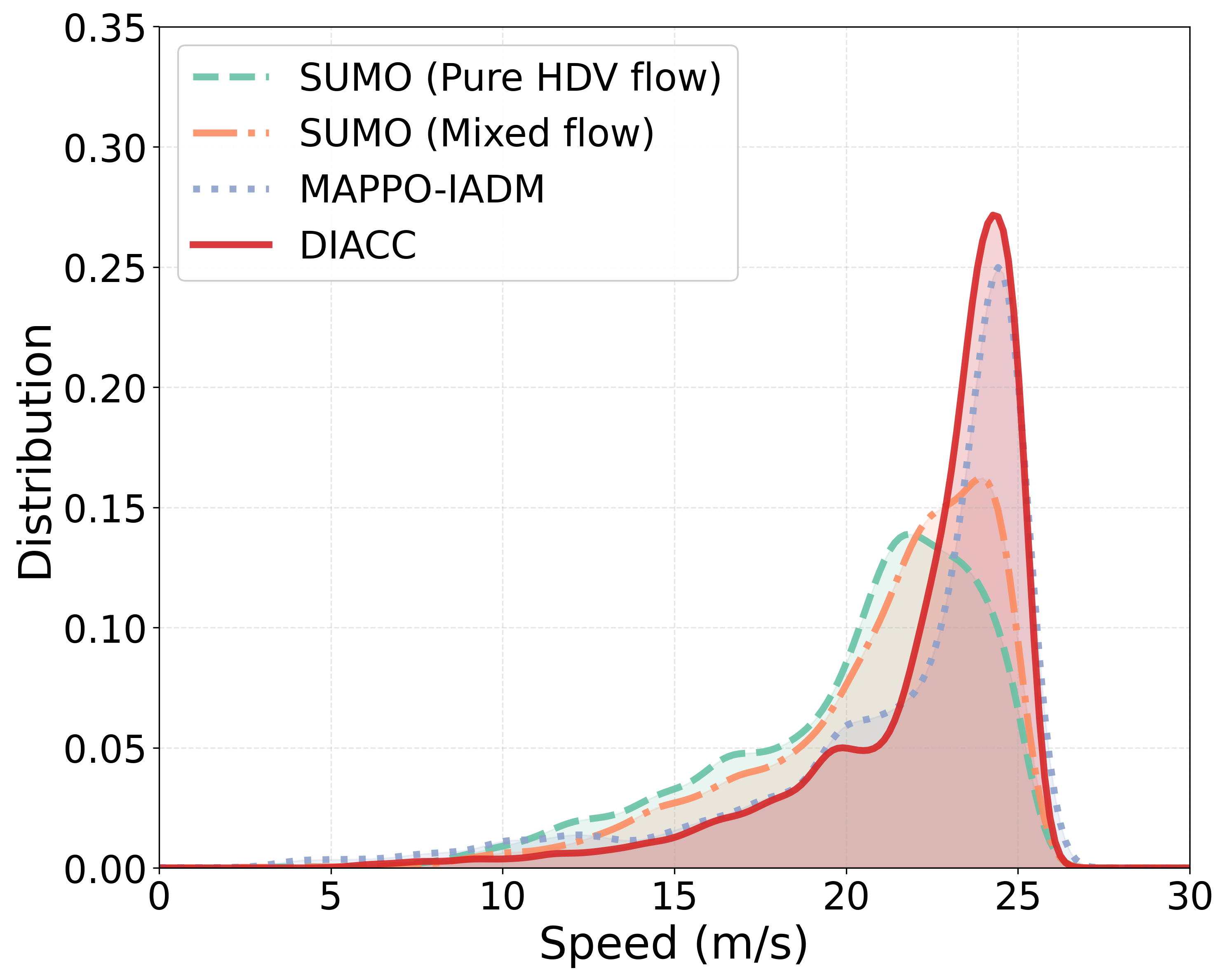}
        \label{fig:speed_kde_N40_normalmultilane}
}
\hfill
\subfloat[$N=40$, 25\% Capacity Reduction]{
        \includegraphics[width=0.31\textwidth]{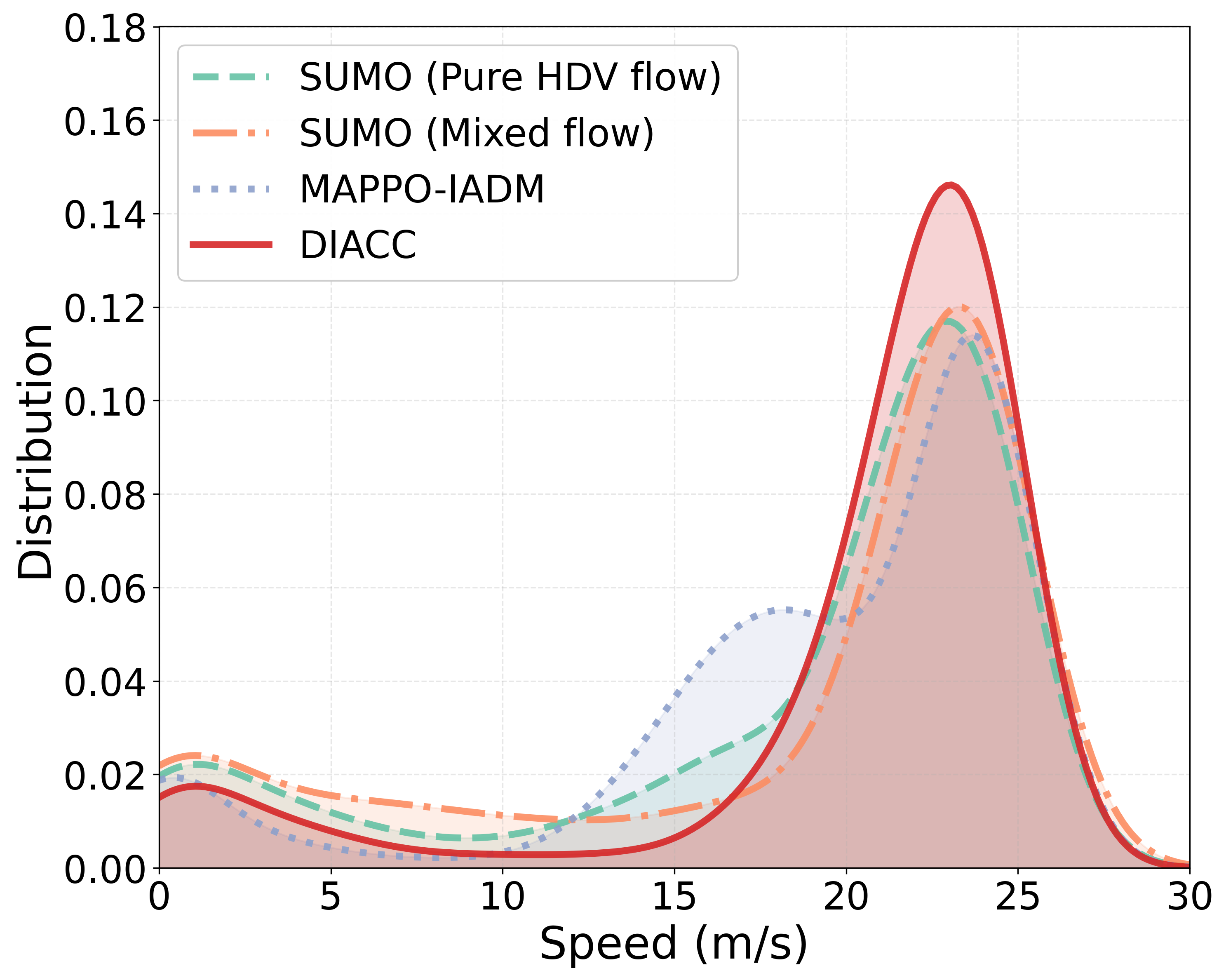}
        \label{fig:speed_kde_N40_25percentreduction}
}
\hfill
\subfloat[$N=40$, 50\% Capacity Reduction]{
        \includegraphics[width=0.31\textwidth]{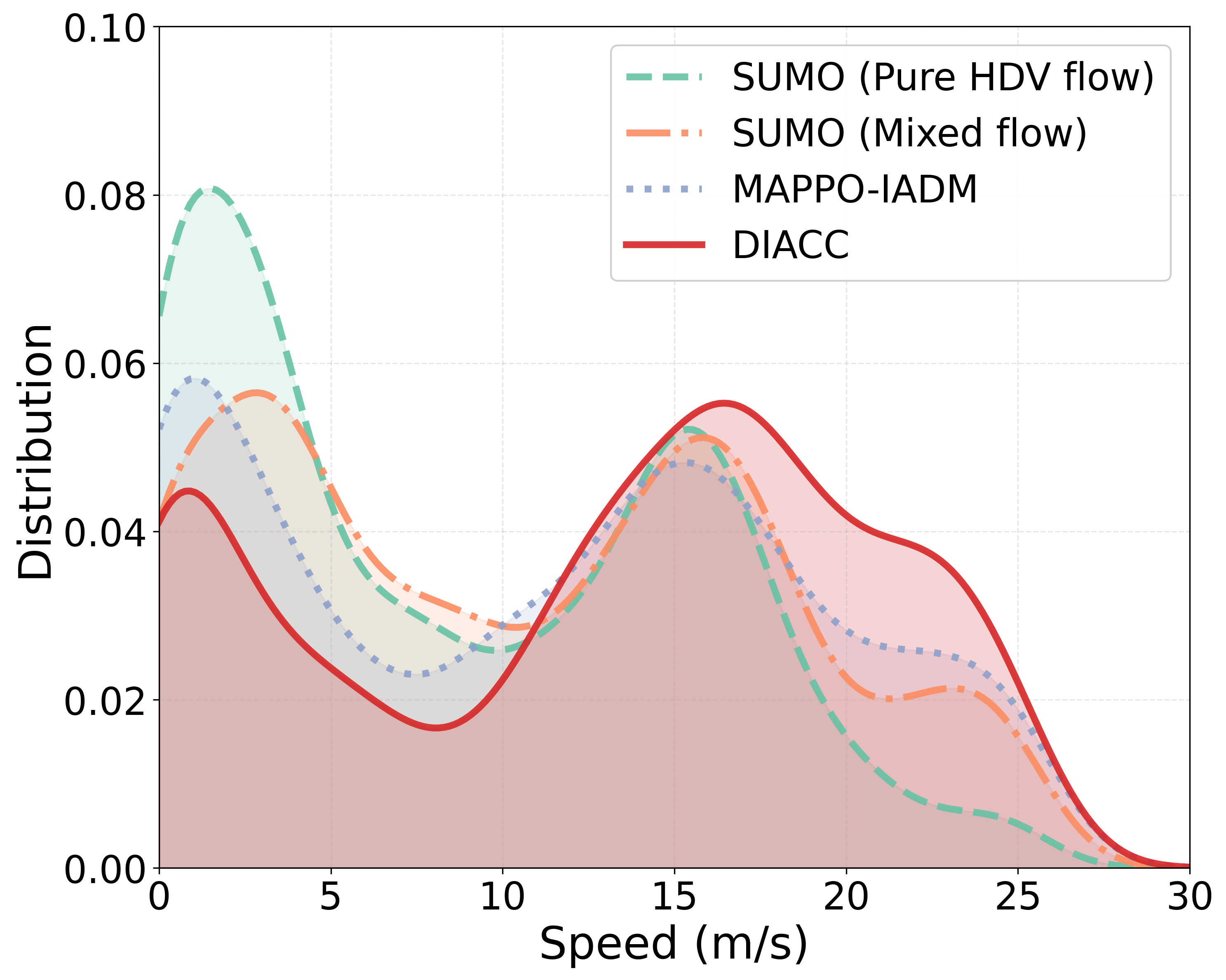}
        \label{fig:speed_kde_N40_50percentreduction}
}
\caption{\color{black}Speed distribution comparison under mixed-traffic scales and capacity-reduction scenarios.
(a)--(b) Box plot and KDE of overall speed distributions.
(c)--(h) KDE of speed distributions across different capacity-reduction scenarios.}
\label{fig:speed_distribution}
\end{figure*}

\vspace{-15pt}

\textcolor{black}{\subsubsection{Cooperation Analysis}
Fig.~\ref{fig:speed_distribution} presents the speed distribution of all methods across traffic scales and capacity-reduction scenarios, 
offering a distributional perspective complementary to the aggregate metrics in Tables~\ref{tab:table_25_reduction} and~\ref{tab:table_50_reduction}.
In the box plot and Kernel Density Estimation (KDE) plots for $N=30$ and $N=40$ (Figs.~\ref{fig:speed_distribution}(a)--(b)), DIACC achieves the highest median speed among all methods, with the most concentrated interquartile range observed under normal multilane and 25\% capacity reduction conditions.
This indicates that DIACC enables vehicles to maintain consistently high and stable speeds, with fewer low-speed oscillations than any baseline.
The KDE plots for individual road segments (Figs.~\ref{fig:speed_distribution}(c)--(h)) further confirm this pattern, 
showing that as capacity reduction increases from 25\% to 50\%, DIACC's speed distribution retains the largest proportion in high-speed regions 
and the smallest in low-speed regions.
This distributional advantage reflects DIACC's dual-level interaction design, where D-IADM equips each actor with proactive interaction-adaptive decisions 
and C-IEC provides accurate global value estimation, jointly sustaining cooperative flow and suppressing speed oscillations as bottleneck severity increases.}
\vspace{-10pt}
\subsection{Discussions}

\textcolor{black}{Vehicle interactions fundamentally influence both individual vehicle movement and overall traffic evolution in mixed traffic environments.
The DIACC strategy addresses this challenge through three complementary components within the MAPPO framework: 
1) D-IADM augments the actor's local interaction perception by distinguishing CAV-CAV cooperative interactions from CAV-HDV observational interactions, 
2) C-IEC strengthens the critic's global traffic understanding through interaction-aware value estimation, and 
3) the cooperative reward mechanism focuses learning on interaction-intensive scenarios through softmin aggregation with temperature annealing.
The experimental results validate the effectiveness of these three contributions.}

\textcolor{black}{The D-IADM module demonstrates its value in enhancing local interaction perception for actor decision-making.
Compared to the vanilla MAPPO baseline, MAPPO-IADM achieves more stable training convergence with lower collision rates, as shown in Figs. \ref{fig:training_all}(a)--(c).
In testing scenarios, MAPPO-IADM reduces the probability of waiting events $p(\mathrm{WEs})$ and safety-critical events $p(\mathrm{SCEs})$ while improving average speeds across various traffic conditions (Tables \ref{tab:table_25_reduction} and \ref{tab:table_50_reduction}).
The differentiated modeling of CAV-CAV and CAV-HDV interactions enables each CAV to better understand the distinct behavioral patterns of surrounding vehicles, leading to more adaptive decision-making in complex traffic environments.}

\textcolor{black}{The C-IEC module validates its contribution in enhancing the critic's global traffic understanding for more effective policy guidance.
The complete DIACC model outperforms MAPPO-IADM particularly in high-density scenarios ($N=30$ and $N=40$), where the advantage becomes more pronounced as traffic complexity increases.
In zero-shot testing conditions, DIACC demonstrates stronger generalization capability compared to MAPPO-IADM, indicating that the interaction-aware value estimation provides more robust guidance for policy learning.
The ITDR module within C-IEC captures how vehicle interactions influence traffic dynamics, enabling the critic to provide more accurate value estimates that guide actor updates toward effective cooperative strategies.}

\textcolor{black}{The reward design mechanism proves essential for learning in interaction-intensive scenarios.
The temperature parameter analysis (Figs.~\ref{fig:training_all}(d)--(f)) shows that fixed $\tau$ values lead to either unstable convergence (small $\tau$) or failure to distinguish scenario difficulty (large $\tau$), while the annealing strategy from $\tau=2$ to $\tau=0.05$ achieves stable training with focused learning on challenging interactions.
The ego reward weight analysis (Figs.~\ref{fig:training_all}(g)--(i)) indicates that $w_e=0.6$ to $0.8$ provides optimal balance between local and global reward signals.
These findings confirm that softmin aggregation effectively directs learning attention toward agents facing complex interaction scenarios, ensuring comprehensive policy improvement across diverse traffic conditions.}

However, there is still considerable room for improvement in the adaptability and robustness of the models in this study. 
In our subsequent work, we plan to:
\begin{itemize}
\item Develop lightweight trajectory prediction modules to enhance the accuracy of safety assessment calculations.
\item Further explore prediction-based MARL models with improved generalization capabilities to better adapt to larger-scale traffic systems.
\item Expand the dual-interaction-aware training framework to diverse traffic scenarios and establish a unified cross-scenario cooperative control framework.
\end{itemize}

\vspace{-12pt}
\section{Conclusion}

\textcolor{black}{Cooperative control of CAVs in mixed traffic environments faces a fundamental tension between decentralized decision-making and global traffic optimization, 
further complicated by the unpredictable behavior of HDVs.
This paper addresses these challenges through the DIACC strategy, which augments the MAPPO framework with dual-interaction-aware capabilities.}

\textcolor{black}{The proposed approach makes three technical contributions that work in concert.
First, the D-IADM module enables each CAV to perceive and respond to the distinct behavioral characteristics of surrounding vehicles by separating the modeling of CAV-CAV collaborative interactions and CAV-HDV adaptive interactions.
Second, the C-IEC module equips the critic with a deeper understanding of how vehicle interactions shape traffic evolution, resulting in more informative value estimates that guide actor policy updates during training.
Third, the reward design mechanism ensures that learning prioritizes scenarios where coordination is most challenging, preventing agents in simpler traffic regions from dominating the training process.}

\textcolor{black}{Experimental validation confirms the effectiveness of these contributions, demonstrating that throughput efficiency is improved under the premise of reduced safety-critical events across diverse traffic conditions.
Furthermore, DIACC maintains this superiority in zero-shot scenarios, demonstrating the generalization capability enabled by interaction-aware learning.}
Future work will focus on extending the theoretical analysis and investigating the scalability of the framework in larger-scale mixed traffic systems and more diverse traffic scenarios.

\color{black}
\section{APPENDIX}
\begin{table*}[htpb]
\centering
\caption{Notations (organized by categories).}
\label{tab:notations}
\begin{tabular}{ll}
\hline
\textbf{Symbol} & \textbf{Definition}\\
\hline
\multicolumn{2}{l}{\textbf{Indexes and Sets}}\\
\hline
$i, j$                 & Vehicle indices (e.g., for CAVs/HDVs) \\
$t$                    & Time step index \\
$l$                    & Lane index \\
$n$                    & Number of CAVs ($|\mathcal{N}|$) \\
$\mathcal{N}$          & Set of CAVs in the mixed traffic \\
$\mathcal{S}$          & Global state space \\
$\mathbf{\mathcal{A}}$ & Joint action space \\
$\mathbf{\mathcal{O}}$ & Joint observation space \\
$\mathcal{M}_{s}^{i}$  & Neighboring vehicles of vehicle $i$ \\
$\mathcal{M}_{w}^{i}$, $\mathcal{M}_{c}^{i}$ & Vehicles within warning / collision ranges \\
$\mathcal{M}_{s,H}^{i}$, $\mathcal{M}_{s,C}^{i}$ & Neighboring HDVs / CAVs \\
$N_{s,H/C}^{i}$          & Number of surrounding HDVs or CAVs of CAV $i$ \\
$\mathcal{G}_{s,H/C}^{i}$ & HDV/CAV interaction graph for CAV $i$ \\
$\mathcal{V}_{s,H/C}^{i}$, $\mathcal{E}_{s,H/C}^{i}$ & Nodes / edges in HDV/CAV interaction graph \\
$\mathcal{G}_{global}$ & Global vehicle interaction graph \\
$\mathcal{V}$, $\mathcal{E}$ & Nodes / edges in global interaction graph \\
\hline
\multicolumn{2}{l}{\textbf{Variables (States, Actions, Observations, etc.)}}\\
\hline
$s_t$, $a_t^i$, $o_t^i$    & Global state; action and observation of vehicle $i$ at time $t$\\
$\tilde{o}_{t+1}^i$ & Augmented observation with $\hat{a}_t^i$ and $a_t^i$ \\
$m_t^i = (x_t^i,y_t^i,v_t^i,\theta_t^i)$ 
& Vehicle $i$'s state (position, speed, heading, etc.)\\
$(\Delta x_t^{i,j}, \Delta y_t^{i,j})$ 
& Relative position of vehicle $j$ w.r.t. vehicle $i$\\
$\Delta v_t^{i,j}$ & Relative speed of vehicle $j$ w.r.t. vehicle $i$\\
$c^j$ & Type of vehicle $j$ (CAV/HDV) \\
$n^l,\;D^l,\;\overline{v}^l,\;\rho^l$ 
& Number of vehicles, density, average speed, and CAV penetration in lane $l$\\
$r_e^i,\;r_g,\;r_{done}$
& Ego reward, global reward, completion reward\\
$r_{e,v}^i,\;r_{e,w}^i,\;r_{e,c}^i,\;r_{e,tp}^i$
& Ego reward components (speed, warning, collision, time) \\
$w^i$
& Softmin weight for agent $i$ \\
$\bar{r}_e$
& Softmin-aggregated ego reward \\
$\mathbf{F}_{t}^{ego},\;\mathbf{F}_{t}^{left},\;\mathbf{F}_{t}^{right},\;\mathbf{F}_{static}$  
& Lane stats and static road structure info\\
$\text{TTC}_t^{i,j},\;d_{t,long}^{i,j},\; \Delta v_{t,long}^{i,j}$
& Time-to-collision, minimum longitudinal head-to-tail distance and longitudinal relative speed\\
$\alpha_{ij}$           
& Attention / importance coefficient between nodes $i$ and $j$\\
$\hat{a}_t^i,\;a_{lat},\;a_{long}$
& Actor command and its lateral / longitudinal components \\
$\mathbf{h}^{H,i}_t,\;\mathbf{f}_t^{H,i},\;\hat{\mathbf{h}}^{H,i}_t$ 
& HDV node features, interaction features, and enhanced node features \\
$\mathbf{f}_t^{C,i}$
& CAV interaction features \\
$\mathbf{f}_t^{dt,i},\;\mathbf{f}_t^{i},\;\mathbf{f}_t^{d}$ 
& Local context, fused observation, global dynamic features \\
$\mathbf{h}^{i}_t,\;\overrightarrow{\mathbf{h}}_{t}^{i}$ 
& Global graph node features and enhanced node features \\
$Q_h,\;K_h,\;V_h$ 
& Query, key, and value in attention mechanism \\
$\overrightarrow{\mathbf{a}},\;\mathbf{\hat{a}}$ 
& Single-layer feedforward networks (used for attention or node update) \\
\hline
\multicolumn{2}{l}{\textbf{Parameters (Constants, Hyperparameters, Functions)}}\\
\hline
$R$                      
& Reward function \\
$P(s_{t+1}\vert s_t,a_t^1,\ldots,a_t^n)$        
& State transition probability \\
$\gamma$                
& Discount factor \\
$\pi_{\theta}$          
& Policy with parameter $\theta$ \\
$J(\pi_{\theta})$       
& Expected cumulative reward \\
$V_\phi$                
& Centralized critic \\
$L$                     
& Horizon length (episode steps) \\
$L^{CLIP},\;L_V,\;V^{CLIP}$ 
& PPO clipped objectives and value prediction \\
$r_t^i(\theta)$         
& Probability ratio (new vs. old policy) \\
$\epsilon$              
& Clipping Hyperparameter (PPO) \\
$\hat{A}_t,\;\delta_t$             
& Advantage estimator and TD error \\
$\lambda$               
& Generalized Advantage (GAE) parameter \\
$T$                     
& Episode length in GAE \\
$w_e,\;w_g$
& Weights for $\bar{r}_e$ and $r_g$ \\
$w_{e,v},\;w_{e,w},\;w_{e,c},\;w_{e,tp}$
& Weights for ego reward components \\
$\tau$
& Temperature parameter for softmin aggregation \\
$\tau_{init},\;\tau_{final}$
& Initial and final temperature for annealing \\
$T_{anneal}$
& Annealing period (half of total training steps) \\
$v_{max}$               
& Prescribed maximum speed \\
$d_{th,w},\;d_{th,c}$    
& Distance thresholds (warning / collision)\\
$v_{th}$                
& Speed threshold for penalty tuning \\
$p_{e,v},\;p_{e,c},\;p_{g,v}$
& Reward bias terms \\
$L_{veh},\;W_{veh}^i$
& Vehicle length and width \\
$S_H,\;S_N$             
& Hidden-layer sizes (e.g., for GAT or MLP) \\
$W,\;W_Q,\;W_K,\;W_V,\;W_O,\;d_k$  
& Attention projection matrices and key dimension \\
$B,\;H$
& Number of attention heads (GAT / cross-attention) \\
$\sigma,\;\sigma(\cdot)$
& Entropy weight and activation function \\
\hline
\end{tabular}
\end{table*}









\bibliographystyle{ieeetr}
\bibliography{ref.bib}

\vskip -1.1cm
\begin{IEEEbiography}[]{Zhengxuan Liu} received the B.E. and M.S. degrees from Tianjin University, 
        Tianjin, China in 2016 and 2019, respectively. 
        From 2019 to 2021, she worked as an EEA Engineer at CATARC (Tianjin) Automotive Engineering Research Institute Co.,Ltd., specializing active safety domain. 
        She is currently pursuing the Ph.D. degree in control science and engineering with the School of Electrical and Information Engineering. 
        Her current research interests include cooperative control and security issue for mixed traffic system.
\end{IEEEbiography}
\vskip -1.12cm
\begin{IEEEbiography}[]{Yuxin Cai} received B.E. degree from Nanyang Technological University (NTU), 
        Singapore in 2022. 
        Now she is a PhD student at School of Mechanical and Aerospace Engineering, NTU, Singapore. 
        Her main research interests are reinforcement learning, multi-robot cooperation and foundation models.
\end{IEEEbiography}
\vskip -1.125cm
\begin{IEEEbiography}[]{Yijing Wang} received her M.S. degree in control
        theory and control engineering from Yanshan University and the Ph.D. degree in control theory from Peking University, China, in 2000 and 2004,
        respectively. 
        In 2004, she joined the School of Electrical and Information Engineering, Tianjin University, where she is a full professor.
        Her research interests are analysis and control of switched/hybrid systems, and robust control.
\end{IEEEbiography}
\vskip -1.125cm
\begin{IEEEbiography}[]{Xiangkun He} 
        (Senior Member, IEEE) is currently a UESTC100 Young Professor at the University of Electronic Science and Technology of China. 
        Previously, he was a Research Fellow at Nanyang Technological University, Singapore, and served as a Senior Research Scientist at Huawei Noah's Ark Lab from 2019 to 2021. 
        He earned his Ph.D. in 2019 from the School of Vehicle and Mobility at Tsinghua University. He has authored over 60 papers and holds 8 granted patents. 
        His research interests include reinforcement learning, trustworthy AI, autonomous vehicles, and robotics. 
        He has received many awards and honors, selectively including the Tsinghua University Outstanding Doctoral Dissertation Award in 2019, 
        Best Paper Finalist at IEEE ICMA 2020, Huawei Major Technological Breakthrough Award in 2021, Best Paper Runner-Up Award at CVCI 2022, 
        and Runner-Up in the Intelligent Algorithm Final of the 2022 Alibaba Global Future Vehicle Challenge. 
        He serves as a reviewer for over 50 renowned journals and conferences.
\end{IEEEbiography}
\vskip -1.125cm
\begin{IEEEbiography}[]{Chen Lv} 
        (Senior Member, IEEE) 
        received the Ph.D. degree from the Department of Automotive Engineering, Tsinghua University, China, in 2016. 
        From 2014 to 2015, he was a joint Ph.D. Researcher with the EECS Department, University of California, Berkeley. 
        He is currently an Associate Professor with Nanyang Technology University, Singapore. 
        His research focuses on advanced vehicles and human–machine systems, where he has contributed over 100 papers and received 12 granted patents in China. 
        He received many awards and honors, including the Highly Commended Paper Award of IMechE U.K. in 2012, the Japan NSK Outstanding Mechanical Engineering Paper Award in 2014, 
        the Tsinghua University Outstanding Doctoral Thesis Award in 2016, the IEEE IV Best Workshop/Special Session Paper Award in 2018, 
        the Automotive Innovation Best Paper Award in 2020, the winner of Waymo Open Dataset Challenges at CVPR 2021, and the Machines Young Investigator Award in 2022. 
        He serves as an Associate Editor for IEEE Transactions on Intelligent Transportation Systems, IEEE Transactions on Vehicular Technology, 
        and IEEE Transactions on Intelligent Vehicles; and a Guest Editor for IEEE Intelligent Transportation Systems Magazine, IEEE/ASME Transactions on Mechatronics, 
        and Applied Energy.
\end{IEEEbiography}
\vskip -1.125cm
\begin{IEEEbiography}[]{Zhiqiang Zuo} 
        (Senior Member, IEEE) 
        received the M.S. degree in control theory and control engineering in 2001 from Yanshan University and the Ph.D. degree in control theory in 2004 from Peking University, China. 
        In 2004, he joined the School of Electrical and Information Engineering, Tianjin University, where he is a full professor. 
        From 2008 to 2010, he was a Research Fellow in the Department of Mathematics, City University of Hong Kong. From 2013 to 2014, he
        was a visiting scholar at the University of California, Riverside. 
        His research interests include nonlinear control, robust control, multi-agent systems and cyber-physical systems.
        
        Dr. Zuo is an associate editor of the Journal of the Franklin Institute (Elsevier).
\end{IEEEbiography}

\end{document}